\documentclass[a4paper,11pt]{article}
\pdfoutput=1 
\usepackage{graphicx}

\usepackage{jcappub} 

\usepackage[T1]{fontenc} 

\title{DBI Galileon inflation in the light of Planck 2015}

\author[a,b]{K. Sravan Kumar}
\author[c,d,e,f]{Juan C. Bueno S\'anchez}
\author[g,h,i]{Celia Escamilla-Rivera}
\author[a,b]{J. Marto}
\author[a,b]{P. Vargas Moniz}

\affiliation[a~]{Departamento de F\'{\i}sica, Universidade da Beira Interior (UBI), \\ 6200, Covilh\~{a}, Portugal}

\affiliation[b~]{Centro de Matem\'atica e Aplica\c{c}\~oes da Universidade da Beira Interior (CMA-UBI), \\ 6200, Covilh\~{a}, Portugal}

\affiliation[c~]{Departamento de F\'{i}sica, Universidad del Valle, Ciudadela Universitaria Melendez, Calle 13 No 100-00, A.A. 25360, Santiago de Cali, Colombia.}

\affiliation[d~]{Centro de Investigaciones en Ciencias B\'asicas y Aplicadas, Universidad
Antonio Nari\~no, Cra 3 Este \# 47A-15, Bogot\'a D.C. 110231, Colombia.}

\affiliation[e~]{Escuela de F\'{i}sica, Universidad Industrial de Santander, Ciudad Universitaria, Cra. 27 Calle 9, Bucaramanga 680002, Colombia.}

\affiliation[f~]{Departamento de F\'{i}sica At\'omica, Molecular y Nuclear, Universidad Complutense de Madrid, Ciudad Universitaria, Plaza Ciencias, 1, 28040, Madrid, Spain.}

\affiliation[g~] {Departamento de F\'{i}sica, Universidade Federal do Esp\'{i}rito Santo,
Av. Fernando Ferrari 514, Vit\'oria, ES, 29075-910, Brazil}

\affiliation[h~]{Mesoamerican Centre for Theoretical Physics (International Centre for Theoretical Physics regional headquarters in Central America, the Caribbean and Mexico),
Universidad Aut\'onoma de Chiapas. Ciudad Universitaria, Carretera Zapata Km. 4,
Real del Bosque (Ter\'an), 29040, Tuxtla Guti\'errez, Chiapas, M\'exico.}

\affiliation[i~]{School of Physics and Astronomy, University of Nottingham, Nottingham NG7 2RD, United Kingdom.}

\emailAdd{sravan@ubi.pt}
\emailAdd{juan.c.bueno@correounivalle.edu.co}
\emailAdd{celia.escamilla@cosmo-ufes.org}
\emailAdd{jmarto@ubi.pt}
\emailAdd{pmoniz@ubi.pt}

\abstract{In this work we consider a DBI Galileon (DBIG) inflationary model
and constrain its parameter space with the Planck 2015 and BICEP2/Keck array and Planck
(BKP) joint analysis data by means of a potential independent analysis. We focus our attention on inflationary solutions characterized by a constant or varying sound speed as well as warp factor. We impose bounds on stringy aspects of the model, such as the warp factor $\left(f\right)$ and the induced gravity parameter $\left(\tilde{m}\right)$. We study the parameter space of the model and find that the tensor-to-scalar ratio can be as low as $r\simeq6\times10^{-4}$ and inflation happens to be at GUT scale. In addition, we obtain the tilt of the tensor power spectrum and test the standard inflationary consistency relation $\left(r=-8n_{t}\right)$ against the latest
bounds from the combined results of BKP+Laser Interferometer Gravitational-Waves Observatory (LIGO), and find that DBIG inflation predicts a red spectral index for the tensor power spectrum.}

\keywords{inflation, string theory and cosmology, primordial gravitational waves (theory)}

\arxivnumber{1504.01348}

\newcommand{\cd}{{\cal D}}
\renewcommand{\[}{\begin{equation}}
\renewcommand{\]}{\end{equation}}

\makeatother

\begin{document}
\maketitle

\section{Introduction}
In the context of the recent results from Planck \cite{Ade:2015ava,Ade:2015lrj,Planck:2015xua},
the joint analysis of BICEP2/Keck Array and Planck (BKP) \cite{Ade:2015tva},
the inflationary paradigm, in particular single-field inflation, seems
to be the one chosen by nature to generate the observed adiabatic,
nearly scale-invariant, Gaussian spectrum of curvature perturbations.
In particular, the Planck satellite, with its exceptional quality
data, constrains the power spectrum tilt of the curvature perturbation
with $n_{s}=0.968\pm0.006$, ruling out scale invariance at more than
$5\sigma$ \cite{Ade:2015lrj}. Moreover, results indicate that we
live in a spatially flat universe, $\Omega_{k}=0.000\pm0.0025$, and
that the perturbation spectrum imprinted in the CMB is Gaussian to
a high degree. This imposes severe constraints on the bispectrum amplitudes:
$f_{NL}^{loc}=0.8\pm5.0$, $f_{NL}^{eq}=-4\pm43$ and $f_{NL}^{ortho}=-26\pm21$
at 68\% confidence level. Furthermore, Planck data \cite{Ade:2015lrj}
suggests a small running $dn_{s}/d\ln{k}=-0.003\pm0.007$ , which
is consistent with the prediction from single-field models of inflation
\cite{Lyth:2009zz}. Although there is no significant detection of
primordial tensor modes, the current analysis points to an upper bound
for tensor-to-scalar ratio $\left(r\right)$ of the order of $\mathcal{O}\left(10^{-2}\right)$.
The proposed post-Planck satellites CMBPol, COrE, Prism, LiteBIRD
and many other ground based experiments such as Keck/BICEP3 \cite{Bouchet:2011ck,Creminelli:2015oda}
are expected to reach enough sensitivity to detect B-modes and further
constrain $r\sim\mathcal{O}\left(10^{-3}\right)$. The latest results
suggest no evidence for a blue tilt of the gravitational wave power
spectra \cite{Ade:2015lrj,Ade:2015tva}. In addition, this result
is supported by the Joint analysis of BKP+Laser Interferometer Gravitational-Waves
Observatory (LIGO) \cite{Huang:2015gka}. According to the new data,
the simplest model with a $\phi^{2}$ potential is ruled out, whereas
Starobinsky's $R^{2}$ inflation and non-minimal coupled models \cite{Starobinsky:1980te,Ade:2015lrj,Boubekeur:2015xza}
remain consistent. In recent works, these two models have been presented
as cosmological attractor models \cite{Galante:2014ifa}. Under the
new constraints on $\left(f_{NL}^{eq}\,,\, f_{NL}^{ortho}\right)$,
inflation with a non-canonical kinetic term has gained importance,
in particular the Dirac-Born-Infeld (DBI) inflation is compatible
with Planck data when the sound speed $c_{s}\geq0.087$ \cite{Ade:2015ava}.
This is interesting in the context of consistent inflationary mechanism
aiming UV-completion theories, such as string theory and supergravity
(SUGRA) \cite{Westphal:2014ana,Baumann:2014nda,Cecotti:2014ipa,Dalianis:2015fpa}.

Our paper provides an observationally consistent inflationary scenario
bearing features of modified gravity and string theory. We focus on
a particular class of string inflation constituting a generic extension
of the DBI inflation \cite{Silverstein:2003hf,Alishahiha:2004eh,
sevaral_DBI1,shaderab,Kobayashi:2007hm,
Baumann:2006cd,Becker:2007ui,
Deser:1998rj,Deffayet:2009mn}.
A peculiar feature of the DBI action is that it involves a non-linear
function of the inflaton's kinetic term, which leads to the generation
of a potentially measurable non-Gaussian signal in the perturbation
spectrum. This is due to a small sound speed which can accommodate
a smaller tensor-to-scalar ratio while sourcing large non-Gaussianities.
However, stringent observational bounds on non-Gaussianity restrict
the validity domain of DBI inflation \cite{Baumann:2006cd,Ade:2015ava,Baumann:2014nda}.
In this paper, we focus on a well-motivated extension of DBI inflation:
the DBI Galileon (DBIG) inflation \cite{deRham:2010eu,Goon:2010xh,Goon:2011qf}.
This model is more generic in the sense that it involves induced gravity,
which arises due to the motion of the D-brane in warped geometry.

DBIG inflation, as it has been generically considered in the literature
\cite{Mizuno:2010ag,RenauxPetel:2011dv,
Gao:2011qe,RenauxPetel:2011uk,Choudhury:2012yh,Koyama:2013wma,Renaux-Petel:2013ppa,Andre:2013afa},
is mainly restricted to the slow-roll regime. Moreover, these studies
are more focussed on the parameter space of the single-field and multifield
DBIG model with respect to the various types of non-Gaussianities.
We must emphasize that in Ref.~\cite{Choudhury:2012yh} a different
motivation is considered, and DBIG inflation is studied in the background
of SUGRA under the assumption of a Coleman-Weinberg type of Galileon
potential. In this paper, we propose to study single-field DBIG inflation
without any particular choice of potential. More precisely, we aim
to constrain the parameter space of the DBIG model according to the
latest Planck 2015 data. We mainly focus our attention in two inflationary
regimes. Namely, those with and without a constant warp factor. We
aim to identify crucial differences between these two scenarios with
respect to the corresponding inflationary predictions. In addition,
in each case, we analyze the deviation from the standard slow-roll
consistency relation $r=-8n_{t}$ due to the effect of induced gravity
on the D-brane. 

The organization of this paper is as follows. In Sec.~\ref{BGS} we briefly describe the model and present the background equations for the DBIG inflation with non-trivial warping \cite{RenauxPetel:2011uk}. In the case of constant sound speed and warp factor, we obtain the exact background solutions. In Sec.~\ref{pert} we compute the amplitude and tilt of the scalar and tensor perturbation spectra for the DBIG model, deferring the details of the computation to Appendix \ref{ap3}. In Sec.~\ref{comp} we study the parameter space of the DBIG model by comparing its predictions in different limits with CMB data. In Sec.~\ref{varysols} we present general background solutions using two different ansatz to integrate analytically the equations of motion. A detailed computation of the approximate solutions can be found in Appendix \ref{ap1}. Finally, we present our conclusions in Sec.~\ref{conclns}.

\section{DBI-Galileon inflationary model}\label{BGS}
We begin by reviewing the DBIG inflationary scenario
following Ref.~\cite{RenauxPetel:2011uk}. Such a setup
considers a D3-brane with tension $T_{3}$ evolving in a ten dimensional
geometry described by the metric, 
\begin{equation}
ds^{2}=h^{-1/2}\left(y^{K}\right)g_{\mu\nu}dx^{\mu}dx^{\nu}+h^{1/2}\left(y^{K}\right)G_{IJ}\left(y^{K}\right)dy^{I}dy^{J}\equiv H_{AB}dY^{A}dY^{B},\label{10Dmetric}
\end{equation}
with coordinates $Y^{A}=\left\{ x^{\mu},y^{I}\right\} $, where $\mu=0,....3$
and $I=1,....,6$. The induced metric on the D3-brane is given by
\begin{equation}
\gamma_{\mu\nu}=H_{AB}\partial_{\mu}Y_{\left(b\right)}^{A}\partial_{\nu}Y_{\left(b\right)}^{B},\label{indmetric10D}
\end{equation}
where the brane is embedded in higher dimensions by means of the functions
$Y_{\left(b\right)}^{A}\left(x^{\mu}\right)$, with the $x^{\mu}$
being the space time coordinates on the brane. In brane inflation,
the role of the inflaton is played by the radial coordinate $\left(\rho\right)$
of the brane that is moving in the extra dimensions. Since we are
only considering single-field inflation in this paper, we choose the
brane embedding as $Y_{\left(b\right)}^{A}\left(x^{\mu}\right)=\left(x^{\mu},\varphi\left(x^{\mu}\right)\right)$.
Then the induced metric can be written as 
\begin{equation}
\gamma_{\mu\nu}=f^{-1/2}\left(g_{\mu\nu}+f\partial_{\mu}\varphi\partial_{\nu}\varphi\right),\label{indmetric5D}
\end{equation}
where $f$ and $\varphi$ are the warp factor and the scalar field
defined by 
\begin{equation}
f=\frac{h}{T_{3}}\,,\quad\varphi=\sqrt{T_{3}}\rho\,.\label{rescale}
\end{equation}
The D3-brane here is embedded in 5D geometry with the induced metric
Eq.(\ref{indmetric5D}). This introduces an additional contribution
in the action known as Galileon term \cite{GC:2009}. The total action
is then given by 
\begin{equation}
S=\int d^{4}x\left[\frac{m_P^{2}}{2}\sqrt{-g}R\left[g\right]+\frac{\tilde{m}^{2}}{2}\sqrt{-\gamma}R\left[\gamma\right]+\sqrt{-g}\mathcal{L}_{brane}\right]\,,\label{actionDBIG}
\end{equation}
where $m_{P}=2.24\times10^{18}$ GeV is the four dimensional reduced
Planck mass, $\tilde{m}$ is a parameter associated with the induced
gravity%
\footnote{$\tilde{m}$ non trivially depends on the warping $h$, see \cite{RenauxPetel:2011uk}.
In this paper, $\tilde{m}$ is treated as a model parameter.%
} and 
\begin{equation}
\mathcal{L}_{brane}=-\frac{1}{f\left(\varphi\right)}\left(\sqrt{\mathcal{D}}-1\right)-V\left(\varphi\right),\label{braneDBI}
\end{equation}
where 
\[
\mathcal{D}\equiv\textrm{det}\left(\delta_{\nu}^{\mu}+f\partial_{\mu}\varphi\partial_{\nu}\varphi\right).
\]
Assuming the flat Friedmann-Lemaître-Robertson-Walker (FLRW) metric
\begin{equation}
ds^{2}=-dt{}^{2}+a^{2}(t)d\boldsymbol{x}^{2}\,.\label{FRW-metric1-1}
\end{equation}
and allowing the warp factor $f$ to vary, the gravitational field
equations for the action in Eq.~(\ref{actionDBIG}) are \cite{RenauxPetel:2011uk}
\begin{equation}\label{Friedmann}
3H^{2}m_{P}^{2}+3\widehat{H}^{2}\frac{\tilde{m}^{2}}{c_{\cd}^{3}}=\frac{1}{f}\left(\frac{1}{c_{\cd}}-1\right)+V\,.
\end{equation}
\begin{equation}\label{Raychaudhuri}
-m_{P}^{2}\dot{H}+\frac{\tilde{m}^{2}H^{2}}{c_{\cd}}\left[-\frac{\dot{\widehat{H}}}{H^{2}}-\frac{c_{\cd}}{h^{1/4}}\left(\frac{h^{1/4}}{c_{\cd}}\right)^{\cdot}\frac{\widehat{H}}{H^{2}}+\frac{3}{2}\left(\frac{1}{c_{\cd}^{2}}-1\right)\frac{\widehat{H}^{2}}{H^{2}}\right]=\frac{\dot{\sigma}^{2}}{2c_{\cd}}\,,
\end{equation}
where $c_{\cd}^{2}\equiv1-f\dot{\sigma}^{2}$ is the squared sound
speed%
\footnote{Note that the sound speed $c_{\cd}$ here depends not only on the
brane dynamics, (as in DBI models \cite{sevaral_DBI1,shaderab,Baumann:2006cd})
but also on the induced gravity \cite{RenauxPetel:2011uk}.}, $\widehat{H}\equiv H-\frac{\dot{f}}{4f}$ and ${\dot{\sigma}}^{2}\equiv G_{IJ}\dot{\phi}^{I}\dot{\phi}^{J}$.
The appearance of Eq.~(\ref{Raychaudhuri}) can be simplified to
\begin{equation}\label{BGS-1}
\dot{H}-\lambda_{1}H^{2}+\lambda_{2}=0
\end{equation}
after introducing the functions 
\begin{eqnarray}
\lambda_{1} & \equiv & \frac{\tilde{m}^{2}}{m_{P}^{2}c_{\cd}+\tilde{m}^{2}}\left[
\frac{\epsilon_f(\eta_f-\epsilon)}4-\frac{d\ln\left(\frac{h^{1/4}}{c_{\cd}}\right)}{d\ln a}\left(1-\frac{\epsilon_{f}}{4}\right)+\frac{3}{2}\left(\frac{1}{c_{\cd}^{2}}-1\right)\left(1-\frac{\epsilon_{f}}{4}\right)^{2}\right]\!\!,\quad\label{lamda1}\\
\lambda_{2} & \equiv & \frac{1-c_{\cd}^{2}}{2f\left(m_{P}^{2}c_{\cd}+\tilde{m}^{2}\right)}\,,\label{lambda2}
\end{eqnarray}
which depend on $\tilde{m}$ and $c_{\cd}$. We also introduce the slow-roll parameters 
\begin{equation}
\epsilon\equiv-\frac{\dot{H}}{H^{2}}\,\,,\,\,\eta\equiv\frac{d\ln\epsilon}{d\ln a}\,\,,\,\,\epsilon_{\cd}\equiv\frac{d\ln c_{\cd}}{d\ln a}\,\,,\,\,\eta_{\cd}\equiv\frac{d\ln\epsilon_{\cd}}{d\ln a}\,\,,\,\,\epsilon_{f}\equiv\frac{d\ln f}{d\ln a}\,\,,\,\,\eta_{f}\equiv\frac{d\ln\epsilon_{f}}{d\ln a}\label{slwrolldef}
\end{equation}
to describe the evolution of the background geometry, the sound speed
and the warp factor. Note also that in the above we take the brane
tension $T_{3}$ to be a constant, as is usually considered.

In the following we obtain solutions to the background equations for the cases when $\lambda_{1,2}$ are constants.

\subsection{Constant sound speed and warp factor}\label{constcase}
\begin{figure}[htbp]
\includegraphics[width=15cm]{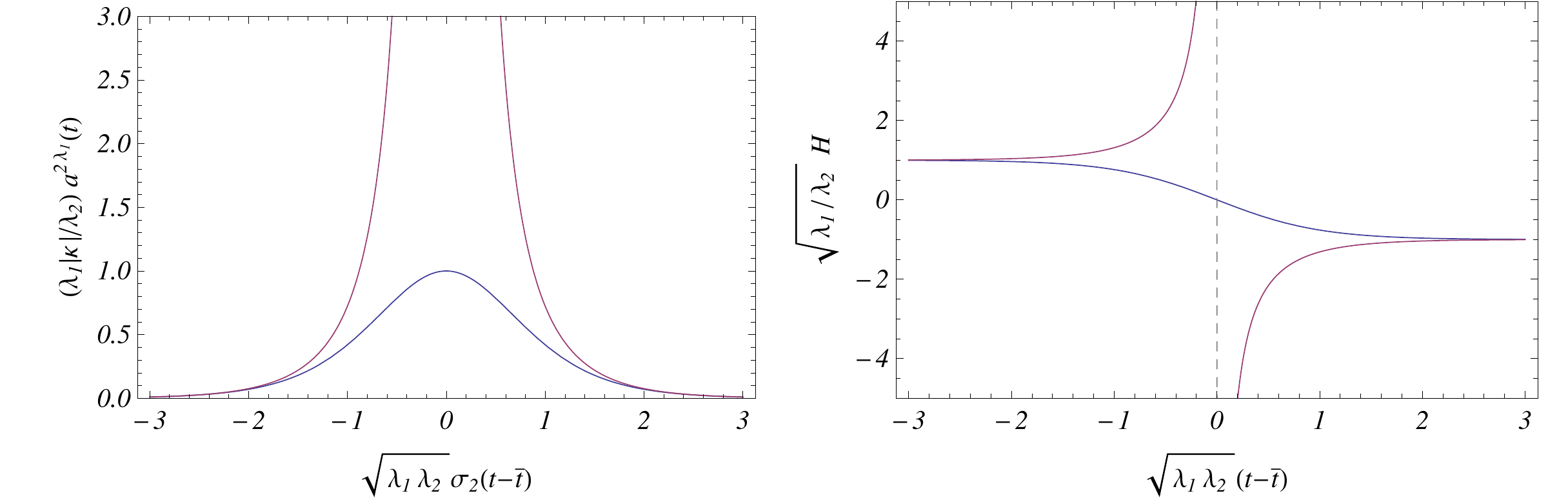}
\caption{Evolution of the scale factor according to Eq.~(\ref{scalefconscase})
(left panel) and the Hubble parameter $H$, according to Eq.~(\ref{Hsolconstcase})
(right panel).}
\label{fig1} 
\end{figure}

Whenever the sound speed $\left(c_\cd\leq1\right)$ and the warp factor are constant, i.e., $\epsilon_\cd=\epsilon_f=0$, the coefficients $\lambda_{1,2}$ in Eq.~(\ref{BGS-1}) are constants. Integrating Eq.~(\ref{BGS-1}) in that case is straightforward. We obtain 
\begin{equation}
H^{2}=\frac{\lambda_{2}}{\lambda_{1}}+\kappa a^{2\lambda_{1}}\,,\label{Hconscase}
\end{equation}
where $\kappa\neq0$ is an arbitrary, dimensionful constant. Writing $H=\dot{a}/a$, the solution to Eq.~(\ref{Hconscase}) is 
\begin{equation}
a^{2\lambda_{1}}(t)=\left(\frac{\lambda_{2}}{\lambda_{1}|\kappa|}\right)\,\exp\left[i\left(1+\sigma_{1}\right)\pi/2\right]\,{\rm sech}^{2}\left[\sqrt{\lambda_{1}\lambda_{2}}\sigma_{2}\left(t-\overline{t}\right)-i\left(1+\sigma_{1}\right)\pi/4\right]\,,\label{scalefconscase}
\end{equation}
where we introduce 
\begin{equation}
\sigma_{1}\equiv{\rm sign}(\kappa)={\rm sign}(\dot{H})\quad,\quad\sigma_{2}\equiv{\rm sign}(\dot{a}).\label{sigma12}
\end{equation}
The explicit time-dependence of the Hubble parameter can be obtained
from Eq.~(\ref{scalefconscase}) 
\begin{equation}
H(t)=-\left(\frac{\lambda_{2}}{\lambda_{1}}\right)^{1/2}\sigma_{2}\tanh\left[\sqrt{\lambda_{1}\lambda_{2}}\sigma_{2}\left(t-\overline{t}\right)-i\left(1+\sigma_{1}\right)\pi/4\right]\,.\label{Hsolconstcase}
\end{equation}
To study inflation we need to set $\sigma_{2}={\rm sign}\left(\dot{a}\right)=+1$,
regardless of $\sigma_{1}={\rm sign}(\dot{H})$. An increasing expansion
rate is obtained for $\sigma_{1}=+1$ ($\lambda_{2}<\lambda_{1}H^{2}$),
which corresponds to the singular behaviour of the scale factor and
the Hubble parameter at $t\to\bar{t}$ (purple line) displayed in
Fig.~\ref{fig1}. A decreasing expansion rate corresponds to $\sigma_{1}=-1$
($\lambda_{2}>\lambda_{1}H^{2}$), in which case both $a(t)$ and
$H(t)$ remain finite throughout the entire evolution (blue line).
In the context of inflation, we focus only on the decreasing expansion
rate $\sigma_{1}=-1$, for which we find a non-singular behaviour
for the scale factor $a(t)$. 

In Sec.~\ref{constcasepert} we impose the necessary conditions to obtain
an inflationary expansion in agreement with current observations.
To do so, in the next section we investigate the scalar and tensor
perturbation spectra, which depend on the slow-roll parameters $\epsilon$
and $\eta$. Using Eq.~(\ref{slwrolldef}) and Eq.~(\ref{Hsolconstcase})
we obtain 
\begin{eqnarray}
\epsilon(t) & = & \lambda_{1}{\rm csch}^{2}\left[\sqrt{\lambda_{1}\lambda_{2}}\sigma_{2}(t-\overline{t})-i(1+\sigma_{1})\pi/4\right]\,,\label{epsconstcase}\\
\eta(t) & = & 2\lambda_{1}{\rm coth}^{2}\left[\sqrt{\lambda_{1}\lambda_{2}}\sigma_{2}(t-\overline{t})-i(1+\sigma_{1})\pi/4\right]\,,\label{etaconstcase}
\end{eqnarray}
from which we arrive at the relations
\begin{equation}\label{relations-eps-eta}
\eta=2\left(\epsilon+\lambda_1\right)\quad,\quad H^2=\lambda_2\left(\lambda_1+\epsilon\right)^{-1}\,,
\end{equation}
where we emphasize that the slow-roll parameter $\eta$ explicitly depends on $\lambda_{1}$. During inflation, $\eta\ll1$ implies
$\lambda_1\ll1$. Therefore, several constraints (to be discussed later on) must be imposed on the model parameters to have $\lambda_{1}\ll1$.

\section{Perturbation spectra}\label{pert}
In this section we portray the generalized approach for calculating the scalar and tensor power spectrum as described
in the context of generalized G-inflation \cite{Kobayashi:2011nu}. We present the calculations of spectral index $\left(n_{s}\right)$,
tensor-to-scalar ratio $\left(r\right)$ and tensor tilt $\left(n_{t}\right)$ up to the slow-roll approximation.

\subsection{Scalar spectrum}
The second order action for scalar perturbations is given by \cite{RenauxPetel:2011uk,Kobayashi:2011nu},
\begin{equation}
S_{s}^{(2)}=\int\, dt\, d^{3}x\left(a^{3}\mathcal{G}_{s}\right)\left[\dot{\zeta}-\frac{\mathcal{F}_{s}/\mathcal{G}_{s}}{a^{2}}\left(\nabla\zeta\right){}^{2}\right]\,,\label{S2pert}
\end{equation}
where $\zeta$ is the curvature perturbation, $t$ is the cosmic time,
$\mathcal{F}_{s}$, $\mathcal{G}_{s}$ are arbitrary functions of
time and \mbox{$c_{s}\equiv(\mathcal{F}_{s}/\mathcal{G}_{s})^{1/2}$}
is the sound speed for scalar perturbations. For the DBIG model \cite{RenauxPetel:2011uk},
the functions $\mathcal{F}_{s}$ and $\mathcal{G}_{s}$ that determine
the second order action for scalar perturbations are 
\begin{equation}
\begin{array}{rcl}
\mathcal{F}_{s}(c_{\mathcal{D}},\epsilon_{\mathcal{D}},\epsilon) & \equiv & B(t)=m_{P}^{2}\left(\epsilon{\cal K}(3{\cal K}-2)+{\cal K}-1\right)+{\displaystyle \frac{\tilde{m}^{2}}{c_{\mathcal{D}}}\left[(\epsilon+\epsilon_{\mathcal{D}}){\cal K}\left(\frac{3{\cal K}}{c_{\mathcal{D}}^{2}}-2\right)+{\cal K}-c_{\mathcal{D}}^{2}\right]}\,,\\
\\
\mathcal{G}_{s}(c_{\mathcal{D}},\epsilon_{\mathcal{D}},\epsilon) & \equiv & A(t)={\displaystyle \frac{m_{P}^{2}}{c_{\mathcal{D}}^{2}}\left(\epsilon{\cal K}^{2}+3c_{\mathcal{D}}^{2}(1-{\cal K}^{2})\right)+{\displaystyle \frac{\tilde{m}^{2}}{c_{\mathcal{D}}^{3}}\left[(\epsilon+\epsilon_{\mathcal{D}}){\cal K}^{2}+3c_{\mathcal{D}}^{2}\left(1-\frac{{\cal K}^{2}}{c_{\mathcal{D}}^{4}}\right)\right]\,,}}
\end{array}\label{FsGs}
\end{equation}
where ${\cal K}\equiv\frac{m_P^{2}+c_{\mathcal{D}}^{-1}\tilde{m}^{2}}{m_P^{2}+c_{\mathcal{D}}^{-3}\tilde{m}^{2}}$.

In addition to the slow-roll parameters previously defined in Eq.~(\ref{slwrolldef}), we introduce the following new parameters 
\begin{equation}
f_{s}\equiv\frac{d\ln\mathcal{F}_{s}}{d\ln a}\quad,\quad f_{s}^{(2)}\equiv\frac{d\ln f_{s}}{d\ln a}\quad,\quad g_{s}\equiv\frac{d\ln\mathcal{G}_{s}}{d\ln a}\quad,\quad g_{s}^{(2)}\equiv\frac{d\ln g_{s}}{d\ln a}\,.\label{ssroll}
\end{equation}
Moreover, using the definition of $c_{s}$ we have 
\begin{equation}
\epsilon_{s}\equiv\frac{d\ln c_{s}}{d\ln a}=\frac{1}{2}\,\left(f_{s}-g_{s}\right)\quad,\quad\eta_{s}\equiv\frac{d\ln\epsilon_{s}}{d\ln a}=\frac{1}{2\epsilon_{s}}\,\left(f_{s}f_{s}^{(2)}-g_{s}g_{s}^{(2)}\right)\quad.\label{sroolscs}
\end{equation}
Although the procedure to obtain the perturbation spectrum is well
known, for the sake of completeness we include in Appendix \ref{ap3}
a detailed computation, obtaining \cite{Kobayashi:2011nu} 
\begin{equation}
{\cal P}_{\zeta}=\frac{\gamma_{s}}{2}\frac{\mathcal{G}_{s*}^{1/2}}{\mathcal{F}_{s*}^{3/2}}\frac{H_{*}^{2}}{4\pi^{2}}\quad,\quad\gamma_{s}\equiv2^{2\nu_{s}-3}\frac{\Gamma\left(\nu_{s}\right){}^{2}}{\Gamma(3/2)^{2}}\left(1-\epsilon_{*}+\frac{g_{s*}}{2}-\frac{f_{s*}}{2}\right)^{2}\,,\label{pwrspectrum}
\end{equation}
where ``$*$'' labels the time of sound horizon crossing when $ky_{s}=-1$. If slow-roll
parameters are sufficiently small one can consider the linear approximation
\cite{Khoury:2008wj,Ribeiro:2012ar}, and then the scalar spectral
index in such case is given by 
\begin{equation}
n_{s}-1\simeq\frac{4\epsilon_{*}+3f_{s*}-g_{s*}}{-2+2\epsilon_{*}+f_{s*}-g_{s*}}\,\,.\label{nsL}
\end{equation}

\subsection{Tensor spectrum}
Similarly to the case of scalar perturbations, the second order action
for tensor perturbations can be written as \cite{Kobayashi:2011nu}
\begin{equation}
S_{t}^{(2)}=\frac{1}{8}\int\, dt\, d^{3}x\left(a^{3}\mathcal{G}_{t}\right)\left[\dot{h}_{ij}^{2}-\frac{\mathcal{F}_{t}/\mathcal{G}_{t}}{a^{2}}(\nabla h_{ij})^{2}\right]\,,\label{st2}
\end{equation}
where $\mathcal{F}_{t}$ and $\mathcal{G}_{t}$ are functions of time
and $c_{t}\equiv(\mathcal{F}_{t}/\mathcal{G}_{t})^{1/2}$ is the sound
speed for tensor perturbations. In the DBIG model, the second order
action is determined by the functions 
\begin{equation}
\mathcal{F}_{t}(c_{\mathcal{D}})\equiv m_{P}^{2}+\tilde{m}^{2}c_{\mathcal{D}}\quad,\quad\mathcal{G}_{t}(c_{\mathcal{D}})\equiv m_{P}^{2}+\frac{\tilde{m}^{2}}{c_{\mathcal{D}}}\,.\label{Fgt}
\end{equation}
Similarly to Eq.~(\ref{ssroll}), we consider now the additional
slow-roll parameters 
\begin{equation}
f_{t}\equiv\frac{d\ln\mathcal{F}_{t}}{d\ln a}\quad,\quad f_{t}^{(2)}\equiv\frac{d\ln f_{t}}{d\ln a}\quad,\quad g_{t}\equiv\frac{d\ln\mathcal{G}_{t}}{d\ln a}\quad,\quad g_{t}^{(2)}\equiv\frac{d\ln g_{t}}{d\ln a}\,,\label{trolls}
\end{equation}
and using the definition of $c_{t}$ we also have 
\begin{equation}
\epsilon_{t}\equiv\frac{d\ln c_{t}}{d\ln a}=\frac{1}{2}\,\left(f_{t}-g_{t}\right)\quad,\quad\eta_{t}\equiv\frac{d\ln\epsilon_{t}}{d\ln a}=\frac{1}{2\epsilon_{t}}\,\left(f_{t}f_{t}^{(2)}-g_{t}g_{t}^{(2)}\right)\,.\label{trollsct}
\end{equation}
It is easy to show now that the tensor spectrum is (see Appendix \ref{ap3})
\begin{equation}
{\cal P}_{t}=8\gamma_{t}\frac{\mathcal{G}_{t*}^{1/2}}{\mathcal{F}_{t*}^{3/2}}\frac{H_{*}^{2}}{4\pi^{2}}\quad,\quad\gamma_{t}\equiv2^{2\nu_{t}-3}\frac{\Gamma\left(\nu_{t}\right){}^{2}}{\Gamma(3/2)^{2}}\left(1-\epsilon_{*}+\frac{g_{t*}}{2}-\frac{f_{t*}}{2}\right)^{2}\,.\label{ptspectrum}
\end{equation}
If slow-roll parameters are sufficiently small one can consider the
linear approximation and obtain the tensor tilt as \cite{Khoury:2008wj,Ribeiro:2012ar}
\begin{equation}
n_{t}\simeq\frac{4\epsilon_{*}+3f_{t*}-g_{t*}}{-2+2\epsilon_{*}+f_{t*}-g_{t*}}\,\,,\label{ntL}
\end{equation}
where the subindex ``$*$'' indicates the time of sound horizon
crossing, determined by the condition $ky_{t}=-1$.

Finally, to further constrain the model parameters with observations we obtain the tensor-to-scalar ratio
\begin{equation}
r\equiv\frac{{\cal P}_{t}}{{\cal P}_{\zeta}}=16\frac{\gamma_{t}}{\gamma_{s}}\left(\frac{\mathcal{G}_{t}}{\mathcal{G}_{s}}\right)^{1/2}\left(\frac{\mathcal{F}_{s}}{\mathcal{F}_{t}}\right)^{3/2}\,.\label{r}
\end{equation}
From Eqs.~(\ref{ntL}) and (\ref{r}) we observe that the standard consistency relation of single-field inflation, $r=-8n_{t}$, is in general violated. In Ref.~\cite{Unnikrishnan:2013rka} it has been shown that, in the case of power law G-inflation, one can have either $r>-8n_{t}$ or $r\leq-8n_{t}$ depending on the model parameters. However, the requirement of subluminal propagation speed of the scalar perturbations restricts $r\leq-\frac{32}{3}n_{t}$.

\section{Comparison to observations}\label{comp}
In this section we study in detail the observational predictions of DBIG inflation and examine the status of the tensor consistency relation. We study the different limits of DBIG inflation beyond the slow-roll approximation and evaluate the effect of higher order corrections in
slow-roll parameters on the model predictions.

We explore the parameter space $\left(c_{\mathcal{D}}\,,\:\tilde{m},\: f\right)$ of DBIG inflation using the Planck constraints on $\left(n_{s},\: r\right)$ and the observed amplitude of the power spectrum $\mathcal{P}_{\zeta_{*}}\simeq2.2\times10^{-9}$ at the pivot scale $k_{*}=0.002\textrm{ Mpc}^{-1}$ \cite{Ade:2015lrj}. In all cases, we find that the predictions of $\left(n_{s},\: r\right)$ do not explicitly depend on the warp factor. Therefore, we first find the range of model parameters $\left(c_{\mathcal{D}},\:\tilde{m}\right)$
compatible with the observed values of $n_{s}=0.968\pm0.006$ and $r<0.1$ at the $95\%$ CL \cite{Ade:2015lrj}. After that, we calculate the tensor tilt $\left(n_{t}\right)$ for the same parameter space that was previously constrained. We expect to
find departures from the consistency relation of single-field inflation, $r=-8n_{t}$. Finally, we compare our results with the BKP+LIGO constraints on the tensor tilt $n_{t}=-0.76_{-0.52}^{+1.37}$ at the $68\%$ CL \cite{Ade:2015lrj,Huang:2015gka}.

\subsection{Constant sound speed and warp factor}\label{constcasepert}
Let us examine the parameter space of DBIG inflation with $\epsilon_\cd=\epsilon_f=0$ in different limits.
For this we use the solutions derived in Sec.~\ref{constcase}. We focus only on the decreasing expansion
rate $\sigma_1=-1$, for which we find a non-singular behaviour for the scale factor $a(t)$.

Firstly, the number of $e$-foldings during inflation can be computed as 
\begin{equation}
N_*=\int_{t_{*}}^{t_{e}}H\, dt\,,\label{efolds}
\end{equation}
where $t_{*}$ is the time when cosmological scales exit the horizon and $t_{e}$ signals the end of inflation, set through the condition
$\epsilon(t_{e})=1$. According to observations, the length of the inflationary phase required to solve the flatness and horizon problems
is around $N_*=40$ to $N_*=70$. Using Eq.~(\ref{Hsolconstcase}) and the condition $\epsilon=1$ to determine $t_{e}$, we integrate
Eq.~(\ref{efolds}) to obtain
\begin{equation}
N_*=\frac{1}{\lambda_{1}}\ln\frac{{\rm cosh}\left[\sqrt{\lambda_{1}\lambda_{2}}\sigma_{2}(t_{*}-\overline{t})-i(1+\sigma_{1})\pi/4\right]}{\sqrt{1+\lambda_{1}}}\,,\label{Nstar}
\end{equation}
which we can relate to the slow-roll parameters $\epsilon$ and $\eta=2\left(\epsilon+\lambda_{1}\right)$ at the time of horizon crossing 
\[
\epsilon_*=\frac{\lambda_{1}}{\left(1+\lambda_{1}\right)\exp[2\lambda_{1}N_*]-1}\,.
\]
Using Eqs.~(\ref{Friedmann}) and (\ref{relations-eps-eta}), we find the scalar potential $V$ in terms of the model parameters
\begin{equation}\label{potentialpdbi}
V=\frac{3\lambda_2}{\lambda_1+\epsilon}\left[m_{P}^{2}+\frac{\tilde{m}^{2}}{c_\cd^3}\right]-\frac{1}{f}\left(\frac{1}{c_{D}}-1\right)\,,
\end{equation}
which allows us to find the energy scale of inflation $V_*^{1/4}$ after evaluating at the time of horizon crossing for cosmological scales. Also, we obtain the mass squared of the inflaton
\begin{equation}
m_{\phi}^{2}=V_{,\phi\phi}=\frac{\ddot{V}}{\dot{\phi}^{2}}\label{mass2phi}\,,
\end{equation}
where 
\begin{equation}
\dot{\phi}^2=\frac{1-c_{\mathcal{D}}^{2}}{f}\label{phiddot}\,.
\end{equation}

\subsubsection{DBI limit: $\tilde{m}\rightarrow0$} \label{dbilimit}
The phenomenology of DBI inflation has been done in recent literature \cite{Weller:2011ey,Li:2013cem} assuming a particular
form of potential. We emphasize here, however, that in our study we do not assume any form of the potential. 

\begin{figure}[htbp]
\centering\includegraphics[width=7.5cm]{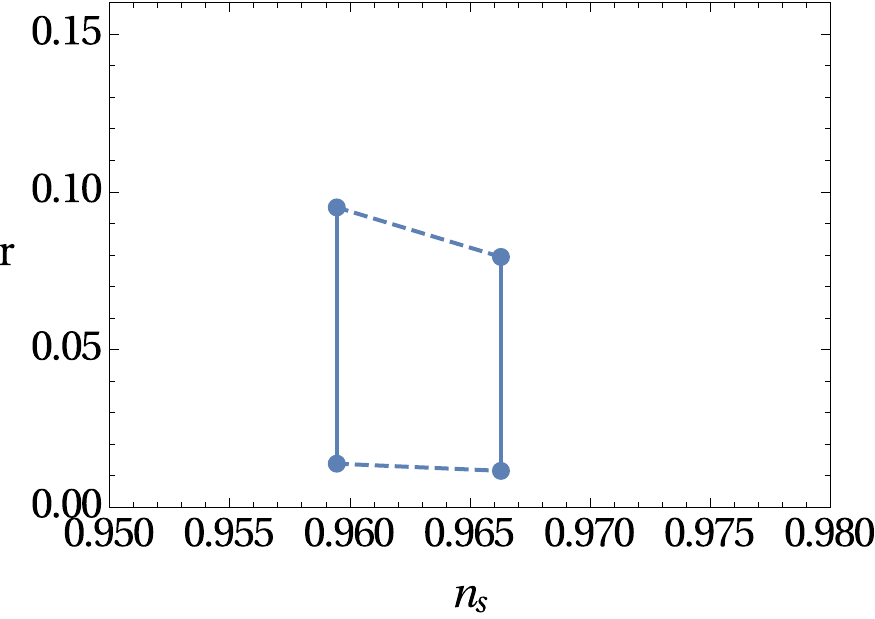}\hspace{0.25cm}\centering
\includegraphics[width=7.5cm]{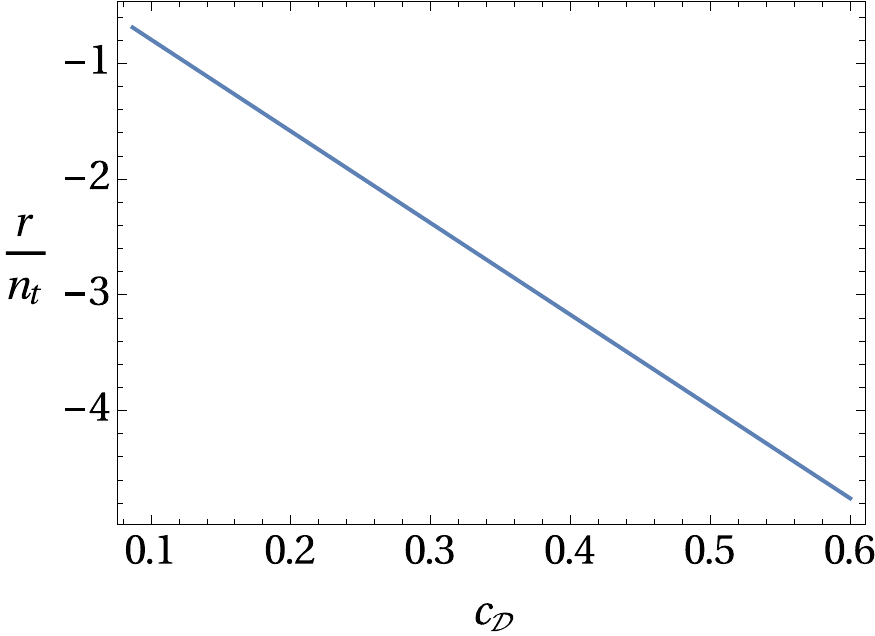}
\caption{In the left panel we depict tensor-to-scalar ratio vs. spectral index where in the plot $N_{*}$ varies from 50 to 60 (from left to right) and $c_\cd$ varies from $0.087$ to $0.6$ (from bottom to top). In the right panel we plot the ratio $r/n_t$ vs. sound speed $c_\cd$ for $N_{*}=60$ .}\label{fig1-1} 
\end{figure}

In this limit $\lambda_{1}\to0$ (see Eq.~(\ref{lamda1})), and we obtain the corresponding background solution from the one obtained
in Sec.~\ref{constcase} as the zeroth order in a series expansion around $\lambda_{1}=0$. Operating similarly for the number of $e$-foldings in Eq.~(\ref{Nstar}) we easily obtain 
\begin{equation}
\lambda_{2}\to\frac{1-c_{\cd}^{2}}{2fm_{P}^{2}c_{\cd}}\quad,\quad H^{2}\to\frac{\lambda_{2}}{\epsilon}\quad,\quad\epsilon\to\frac{1}{1+2N}\quad,\quad\eta\to2\epsilon\quad,\quad{\cal P}_{\zeta}\to\frac{H^{2}}{8\pi^{2}\epsilon c_{\cd}}\,,\label{DBIlargeN}
\end{equation}
where $N$ is the number of e-foldings before the end of inflation.
Fixing the number of $e$-foldings and the amplitude of the perturbation
spectrum we constrain the warp factor $f$. Since we treat $c_{\cd}$
as a model parameter, we obtain its range from the prediction for
non-Gaussianity $f_{{\rm NL}}^{eq}=-\frac{35}{108}\left(\frac{1}{c_{\mathcal{D}}^{2}}-1\right)$
in DBI models \cite{Silverstein:2003hf,Alishahiha:2004eh}. Although
more accurate expressions exist in the literature \cite{Mizuno:2010ag,
RenauxPetel:2011dv,Gao:2011qe,
RenauxPetel:2011uk,Choudhury:2015yna,
Koyama:2013wma,Renaux-Petel:2013ppa},
for our purposes it suffices to consider this simple estimate. This
is appropriate since in the absence of a clear detection of non-Gaussianity
\cite{Ade:2015ava}, the use of more elaborate or complicated expressions
is, in principle, uncalled for. Therefore, in this paper we will not
be concerned with non-Gaussian computations and will use the above
expression to constrain the sound speed $c_{\mathcal{D}}$. The analysis
of the Planck data on $r<0.1$ and $f_{NL}^{eq}=-4\pm43$ allows to
set a conservative bound for this $0.087\leq c_{\cd}\leq0.6$ \cite{Ade:2015lrj,Ade:2015ava}.
Note that larger values of $c_{\cd}$, albeit allowed by the bound
from non-Gaussianity, are disfavoured as they result in a tensor-to-scalar
ratio in excess of the current bound $r<0.1$. Fig.~\ref{fig1-1}
represents the viability of the DBI model. Because of the stringent
bound on $f_{NL}^{eq}$ the DBI inflation is not capable to induce
$r<0.01$ which is consistent with previous studies \cite{Baumann:2006cd,Peiris:2007gz}.
The range of model parameters obtained for $0.087\leq c_{\cd}\leq0.6$
can be found in Table~\ref{summary}. In Fig.~\ref{fig1-1} we depict
our results in the DBI limit. 

\subsubsection{Galileon limit: $\tilde{m}\gg m_{P}$}\label{glimit-1}
Although studying this limit is not generic with
respect to the structure of DBIG, this would nevertheless be useful
to understand the role of induced gravity. Since $c_{\cd}\lesssim1$,
Eq.~(\ref{BGS-1}) gives 
\begin{equation}
\lambda_{1}=\frac{3}{2}\left(\frac{1}{c_{\mathcal{D}}^{2}}-1\right)\quad,\quad\lambda_{2}\equiv\frac{1-c_{\cd}^{2}}{2c_{\cd}f\tilde{m}^{2}}\,.\label{lambdasdbi-1}
\end{equation}
The slow-roll parameters in this case which are given below 
\begin{equation}\label{swrll-G-inf}
\epsilon=\frac{3\left(1-c_\cd^2\right)}{\left(3-c_\cd^2\right)e^{\frac{3}{2}\left(\frac{1}{c_{\mathcal{D}}^{2}}-1\right)N}-2c_\cd^2}\quad,\quad
\eta=3\left(\frac{1}{c_{\mathcal{D}}^{2}}-1\right)+2\epsilon\,.
\end{equation}

\begin{figure}[htbp]
\centering\includegraphics[width=6.5cm]{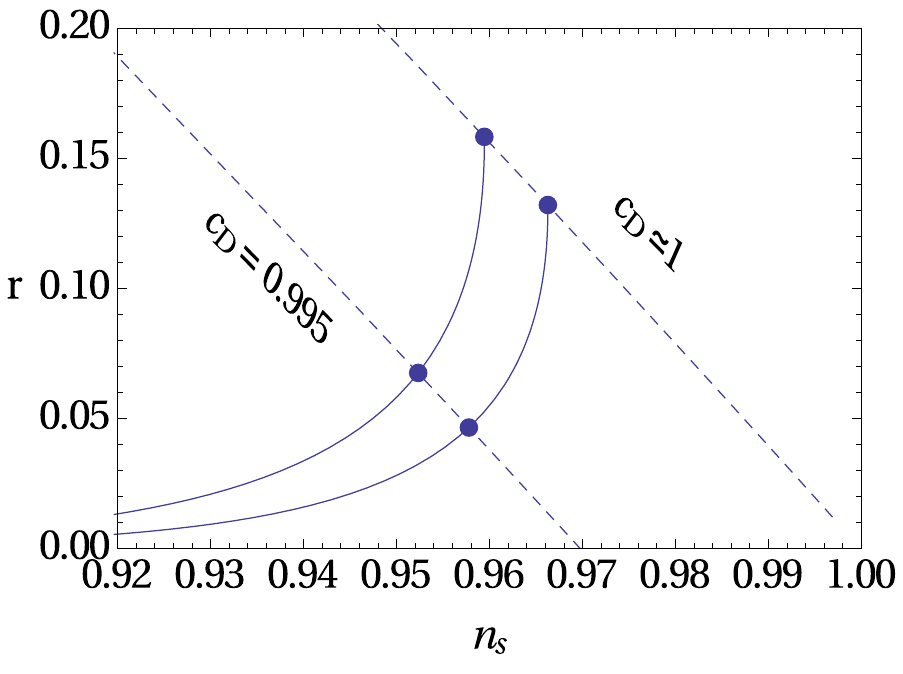}\hspace{0.5cm} \centering
\includegraphics[width=7.0cm]{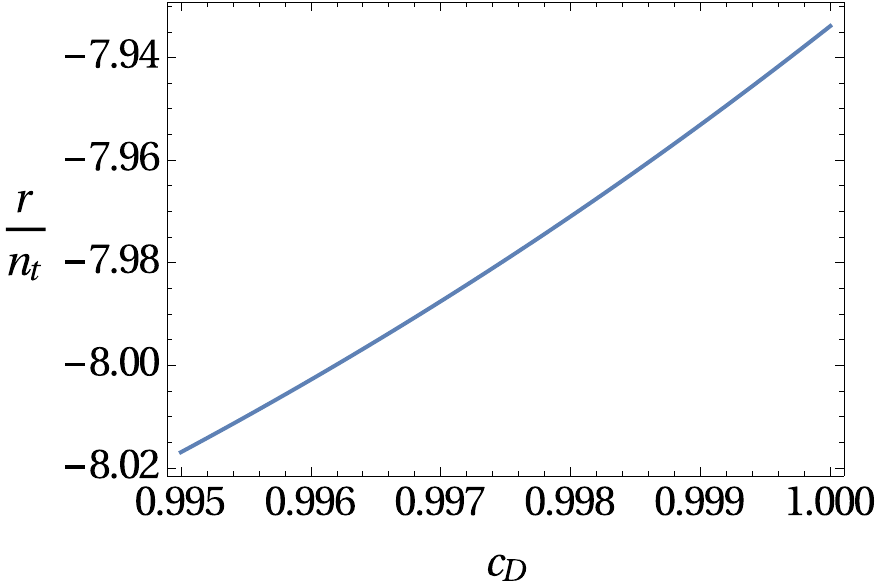}
\caption{Plots of spectral index $n_{s}$ vs. tensor-to-scalar ratio $r$ (left)
and the ratio $r/n_t$ vs. sound speed $c_\cd$ (right)
in the Galileon limit. In the left panel, we take $N_{*}$ varying
from 50 to 60 (from bottom to top). For the right panel we considered $ N_{*}=60 $.}\label{glimit} 
\end{figure}

Unlike in the DBI limit (cf. Eq.~(\ref{DBIlargeN})), in the Galileon
limit, the slow-roll parameters explicitly depend on the sound speed.
It is obvious from Eq.~(\ref{swrll-G-inf}) that $c_{\mathcal{D}}\ll1$
would actually spoil the smallness of $\eta$. Therefore, in this
case we need to keep the sound speed in the narrow range $0.995\leq c_{\cd}<1$
for the results to agree with the current Planck data. Any value of
$c_{\mathcal{D}}<0.995$ would essentially spoil the prediction of
the spectral index and its value would be significantly
out of the current bounds $n_s=0.968\pm0.006$. Therefore, observationally viable inflation due to the induced gravity
term sets $c_{\cd}\lesssim1$, thus resulting in small non-Gaussianities.
This allows to discriminate between the current case and the DBI limit
previously studied. Also, the consistency of the predictions with
data becomes better as the number of $e$-foldings reduces. In particular,
for $N_*\sim50$ our results are perfectly consistent with current
data whereas for $N_*\gtrsim60$ the model is ruled out. Our results
in this case are depicted in Fig.~\ref{glimit}. The derived model
parameters for $0.995\leq c_{\cd}\leq1$ can be found in Table~\ref{summary}. 

\subsubsection{DBI-Galileon case}\label{intercomp}
In this section we consider Einstein and Galileon gravity are on an equal footing. In this case 
\[
\lambda_{1}\equiv\frac{\tilde{m}^{2}}{m_{P}^{2}c_{\cd}+\tilde{m}^{2}}\left[\frac{3}{2}\left(\frac{1}{c_{\cd}^{2}}-1\right)\right]\quad,\quad\lambda_{2}\equiv\frac{1-c_{\cd}^{2}}{2f\left(m_{P}^{2}c_{\cd}+\tilde{m}^{2}\right)}\,.
\]
The corresponding slow-roll parameters are (expressing in the units of $ m_{P}=1 $)
\begin{equation}\label{eps-DBIG}
\epsilon=\frac{3\left(1-c_{\mathcal{D}}^{2}\right)\tilde{m}^{2}}{\left[2c_{\mathcal{D}}^{3}-\left(c_{\mathcal{D}}^{2}-3\right)\tilde{m}^{2}\right]e^{\frac{3\left(\frac{1}{c_{\mathcal{D}}^{2}}-1\right)\tilde{m}^{2}N}{c_{\mathcal{D}}+\tilde{m}^{2}}}-2c_{\mathcal{D}}^{2}\left(c_{\mathcal{D}}+\tilde{m}^{2}\right)}\,\,,\,\,\eta= \frac{3\left(\frac{1}{c_{\mathcal{D}}^{2}}-1\right)\tilde{m}^{2}}{\left(c_{\mathcal{D}}+\tilde{m}^{2}\right)}+2\epsilon\,.
\end{equation}

\begin{figure}[htbp]
\centering\includegraphics[width=9cm]{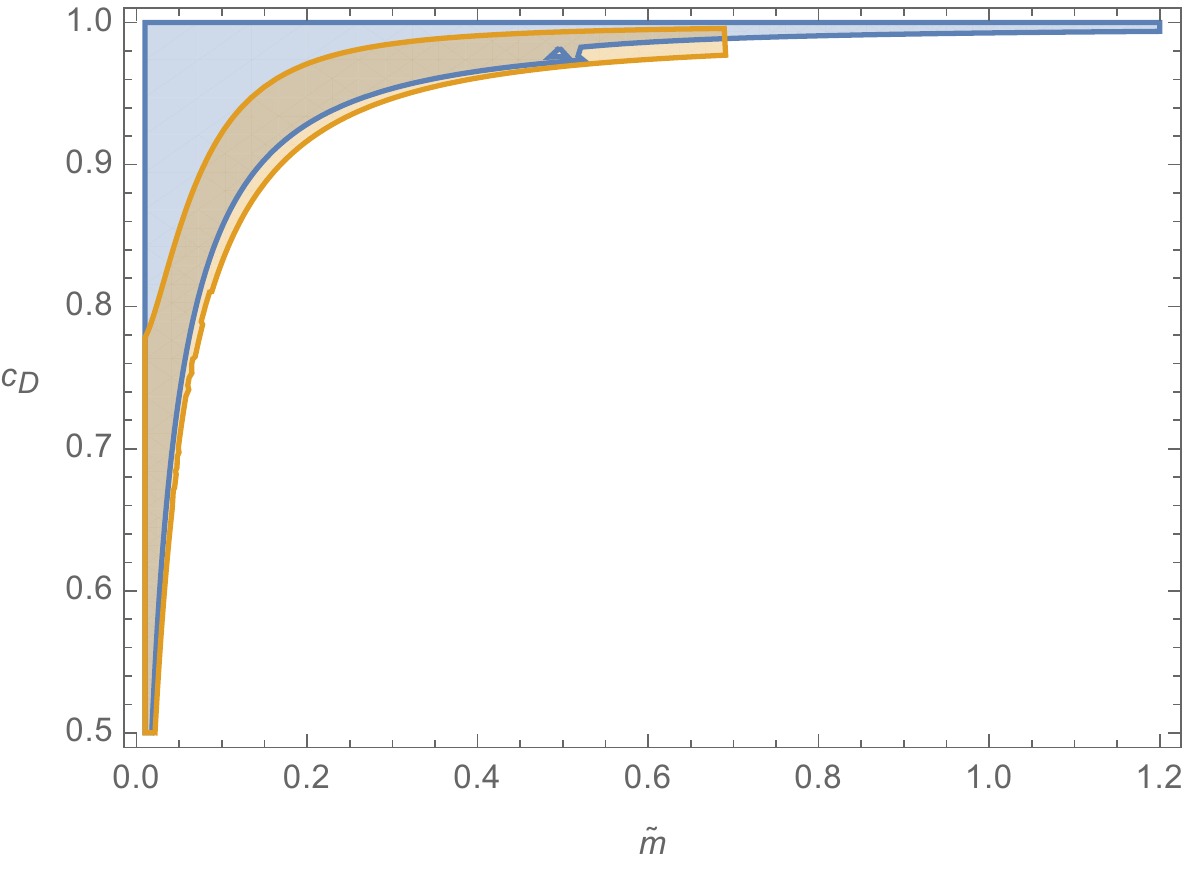}\caption{Contour plots in the plane $(\tilde m,c_\cd)$ (with $\tilde m$ in units of $m_P$). Blue and orange regions represent the space where $n_s=0.968\pm0.006$ and $0.01\leq r\leq0.1$, respectively.}\label{PspaceDBIG} 
\end{figure}

Similarly to the Galileon limit studied in Sec.~\ref{glimit-1}, the sound speed needs to be tuned to $c_{\cd}\simeq0.98-0.99$ to keep the slow-roll parameter $\eta$ small enough to have $n_s=0.968\pm0.006$. We find that $c_{\cd}<0.98$ would essentially spoil the prediction of scalar tilt. We also note here that if $c_{\mathcal{D}}=1$ we obtain exact scale invariance, i.e. $n_s=1$. Since the slow-roll parameter $\epsilon$ in Eq.~(\ref{eps-DBIG}) depends on the parameter $\tilde{m}$, the tensor-to-scalar ratio varies for different values of the induced gravity parameter $\tilde{m}$. This allows us to identify the range of the parameters consistent with current data. In Fig.~\ref{PspaceDBIG} we study the parameter space $\left(c_{\mathcal{D}}\,,\,\tilde{m}\right)$ using the bounds on $\left(n_{s}\,,\, r\right)$. The plot shows that, in the limit $\tilde{m}\to0$, the model reduces to DBI case. Moreover, unless $\tilde m<m_P$, the effect of the induced gravity forces us to constrain the sound speed to $c_{\cd}\sim1$ in order to maintain the agreement with observations.

To constrain the model parameters $\left(c_{\mathcal{D}}\,,\,\tilde{m}\right)$
with the bounds of $\left(n_{s}\,,\, r\right)$ it is also necessary
to check if non-Gaussianities are large. Since the full study of
non-Gaussianity is beyond the scope of this paper, we use the results
in Ref.~\cite{RenauxPetel:2011uk}, where the authors study non-Gaussianity
in the multifield DBIG inflation model. We adopt their expression for $f_{NL}^{eq}$
in the single-field limit, i.e. taking the adiabatic and isocurvature
mode transfer function $T_{\sigma s}\rightarrow0$. We thus constrain
our parameter space using the approximate expression \cite{RenauxPetel:2011uk}
\begin{equation}\label{fnleinter}
f_{NL}^{eq}=-\frac{5}{324c_{\cd}^{2}}\frac{21-404\alpha+2233\alpha^{2}-3066\alpha^{3}}{(1-5\alpha)^{2}(1-9\alpha)}\,\,,\,\,\alpha\equiv\frac{fH^{2}\tilde{m}^{2}}{c_{\mathcal{D}}^{2}}\,.
\end{equation}

\begin{figure}[htbp]
\centering\includegraphics[width=7.0cm]{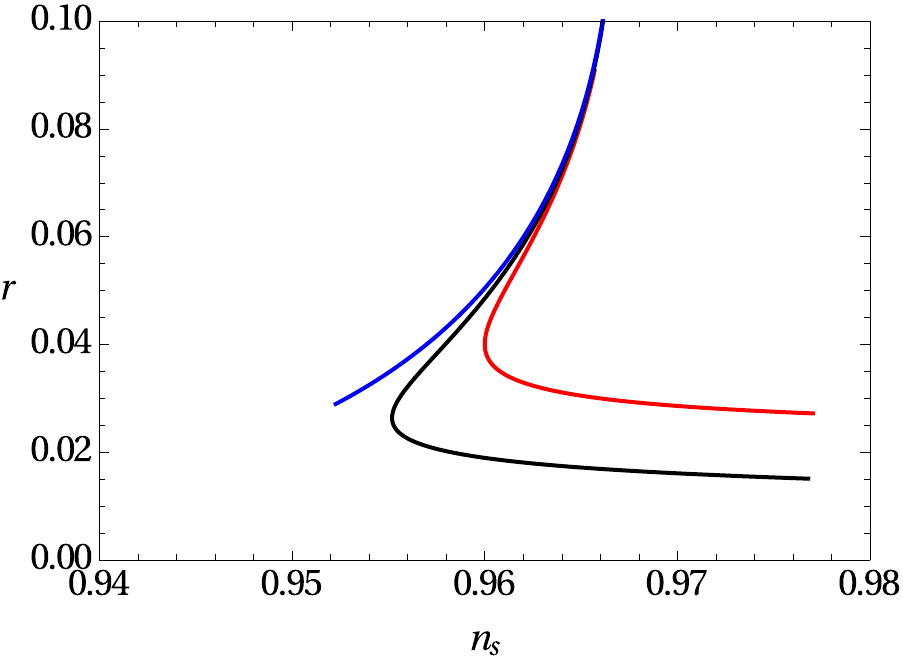}\hspace{0.5cm} \includegraphics[width=7.0cm]{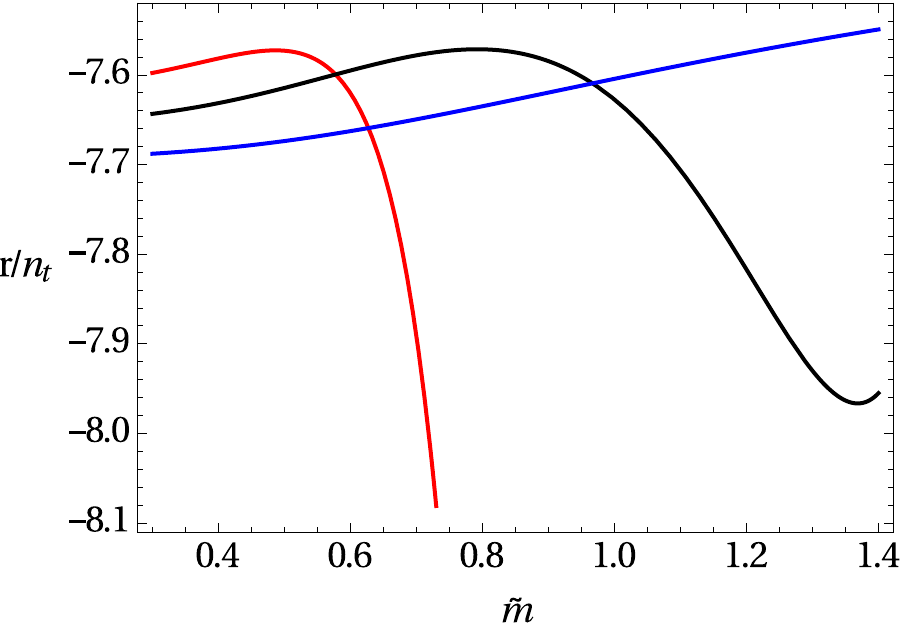}
\caption{Plots of spectral index $n_s$ vs. tensor-to-scalar ratio $r$ (left
panel) and the ratio $r/n_t$ vs. $\tilde{m}$ (with $\tilde m$ in units of $m_P$) (right panel) in the DBIG model. In the left panel we take $c_\cd=0.98$ and $0.3\leq\tilde{m}/m_P\leq0.72$ (red), $c_{\cd}=0.985$ and $0.5\leq\tilde{m}/m_P\leq1.25$ (black), $c_{\cd}=0.99$ and $0.5\leq\tilde{m}/m_P\leq1.25$ (blue). In the plotted curves $\tilde m$ increases as $r$ decreases. In the right panel, the plotted curves correspond to $c_\cd=0.98$ (red), $c_\cd=0.985$ (black) and $c_\cd=0.99$ (blue).}\label{intercase} 
\end{figure}

Setting $N_*=60$, in Fig.~\ref{intercase} we plot the model predictions in the plane $(n_s,r)$ (left panel) for different values of $c_\cd$ and for different ranges of $\tilde m$, as indicated. In the plotted curves, the tensor-to-scalar ratio decreases as we increase $\tilde m$. Therefore, our results show that an increase of the induced gravity lowers the tensor-to-scalar ratio. In the right panel we plot the ratio $r/n_t$ as a function of $\tilde m$. In the range of values of $c_\cd$ consistent with the observed value of the spectral index we find a slight deviation from the standard consistency relation. Nevertheless, such a deviation does not seem to be sufficiently significant to be detected with confidence.

\begin{figure}[htbp]
\centering\includegraphics[width=7cm]{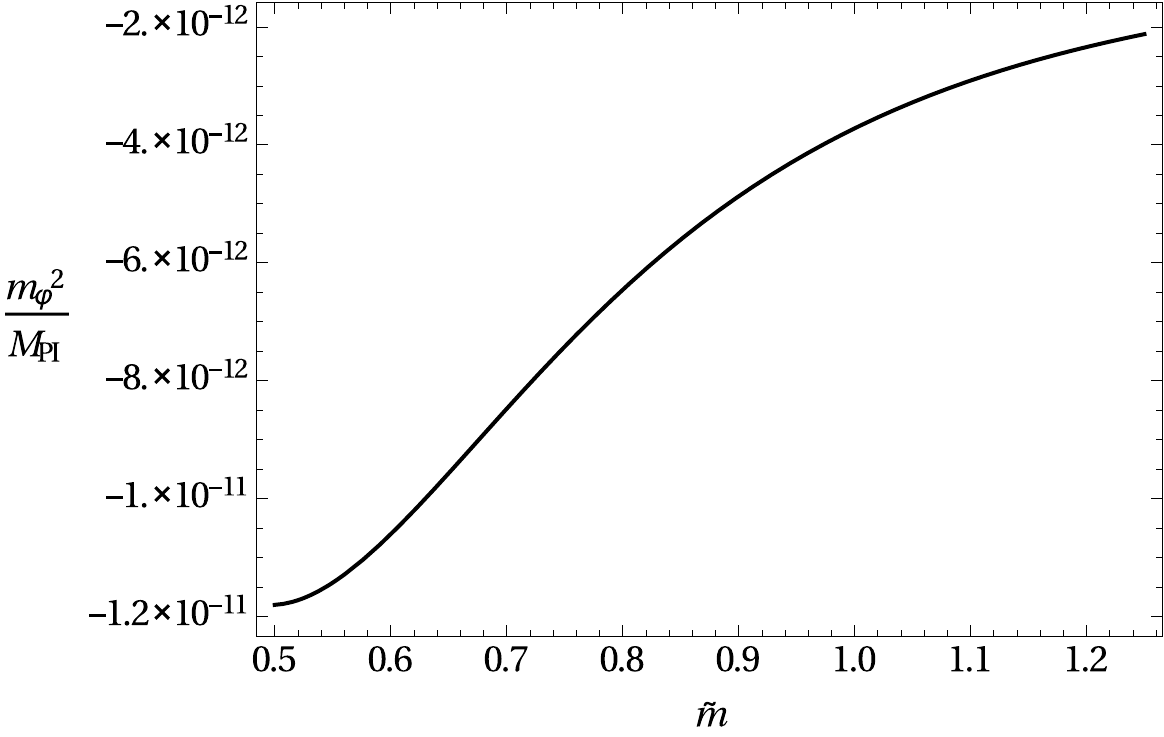}\hspace{0.5cm} \includegraphics[width=6.5cm]{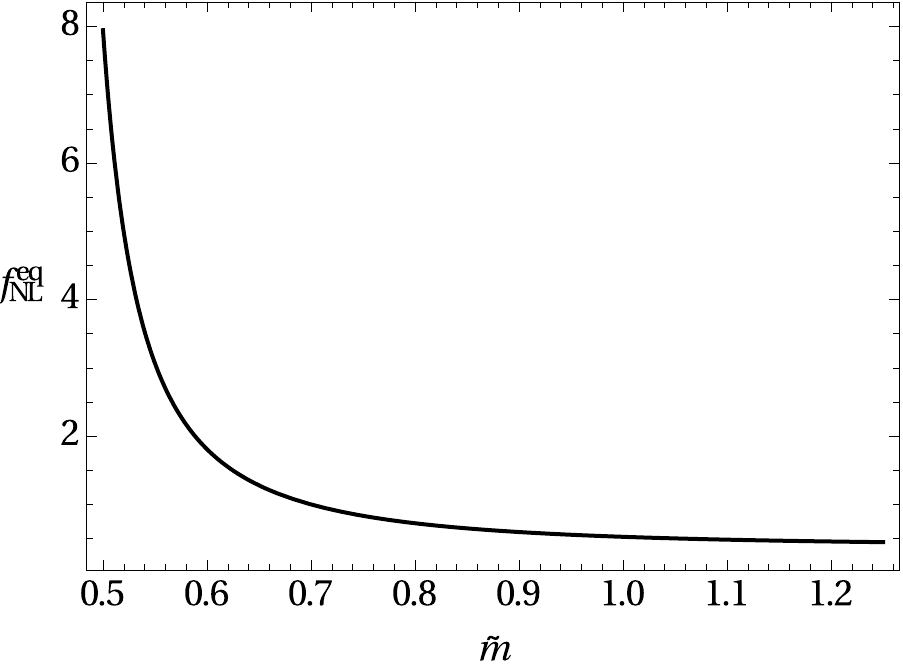}
\caption{Plots of the mass squared of the inflaton field (left panel) and the non-Gaussian parameter $f_{NL}^{eq}$ (right panel) as a function of $\tilde{m}$ (with $\tilde m$ in units of $m_P$). In this plot $ 0.22\leq\alpha\leq0.32 $ for $ 0.5\leq\tilde{m}\leq1.25 $. We take $c_{\cd}=0.985$ to build the plots, hence the depicted behaviour corresponds to the black line in Fig.~\ref{intercase}.}\label{fnl} 
\end{figure}

In Fig.~\ref{fnl} we plot the mass squared of the inflaton, as obtained from Eq.~(\ref{mass2phi}) evaluated at the time of horizon crossing for cosmological scales (left panel), and $f_{NL}^{eq}$ calculated from Eq.~(\ref{fnleinter}) (right panel). From the left plot, we find that the inflaton is tachyonic, whereas for smaller values of $\tilde{m}$, we recover a potential with positive curvature, in agreement with the DBI case. In this sense, it may be worth mentioning that the authors in   Ref.~\cite{Bernardini:2013tba} have studied the possibility that the Born-Infeld tachyon be equivalent to a scalar field in an effective field theory in different warped geometries. Moreover, in Ref.~\cite{Li:2013cem} the observational constraints on tachyon and DBI inflation were studied, and the authors showed that tachyon inflation fits better with cosmological data than DBI. It is also important to notice that $n_t<0$ in all cases, which is statistically preferred by data after the Planck and BKP joint analysis \cite{Ade:2015lrj,Ade:2015tva}, Also, the joint analysis of BKP+LIGO indicates a red tensor tilt $n_{t}=-0.76_{-0.52}^{+1.37}$ at the $68\%$ CL \cite{Huang:2015gka}. In Table~\ref{summary} we report the values of the ratio $r/n_t$, which only results in a slight deviation from the standard consistency relation in most of the cases. We recall that future cosmology probes will be able to discriminate inflationary models by direct detection of primordial B-modes \cite{Creminelli:2015oda}. Finally, from the right panel of Fig.~\ref{fnl} we find that the non-Gaussianity parameter $f_{NL}^{eq}$ is consistent with the stringent bounds imposed by Planck data \cite{Ade:2015ava}.

\subsection{Varying both sound speed and warp factor}\label{varycdf}
The cases considered in Sec.~\ref{constcasepert} (constant sound speed and constant warp factor) are consistent with observational data. However, it is interesting to understand the cases with varying $c_{\cd}$ and $f$. The questions we can pose in these cases are, can we get a parameter space with $r\sim\mathcal{O}\left(10^{-3}\right)$? How do the warped geometries and the scale of inflation change when $\left(c_{\cd},f\right)$ change with time? What is the nature of inflaton field is such cases? In this section, we obtain exact background solutions in two cases: a slowly varying sound speed at fixed warp factor and a slowly varying warp factor at fixed sound speed.

\subsubsection{Varying sound speed ($\epsilon_\cd\neq0,\eta_\cd=0$) and constant warp factor ($\epsilon_f=0$)}\label{varycd}
We assume a slow variation of the sound speed, i.e. $\epsilon_{\mathcal{D}}\ll1$. Using the definition of slow-roll parameters
from Eq.~(\ref{slwrolldef}), we can approximate $c_{\cd}$ in terms of $N=\ln a$ as 
\begin{equation}
c_{\mathcal{D}}=c_{d}\exp\left(\epsilon_{\mathcal{D}}N\right)\simeq c_{d}\left(1+\epsilon_{\mathcal{D}}N\right),\label{cdslwvary}
\end{equation}
where $c_{d}$ is a constant whose magnitude is set some four e-foldings after the largest cosmological scales exit the horizon.

To integrate the background Eq.~(\ref{BGS-1}) it is now convenient to rewrite it as 
\begin{equation}\label{varyspeddBG}
H^{\prime}-\lambda_{1}H+\frac{\lambda_{2}}{H}=0\,,
\end{equation}
where $\lambda_{1,2}$ are computed using the approximation in Eq.~(\ref{cdslwvary}) and the prime stands for $'\equiv\frac{d}{dN}$. Integrating Eq.~(\ref{varyspeddBG}) we obtain the solution $H=H(N)$. To fix the integration constant in the solution it suffices to impose that $\epsilon\equiv-\frac{H'}{H}=1$ at the end of inflation. We choose not to include here the solution $H=H(N)$ as it is a complicated expression involving imaginary error functions \cite{Abramowitz}. To constrain the model parameters we proceed as in Sec.~\ref{constcasepert}. Since in this case $\left(n_{s}\,,\, r\right)$ do not depend on warp factor $f$, we may find the range for $\left(c_{d}\,,\,\tilde{m\,},\,\epsilon_{\cd}\right)$ using the current bounds on $\left(n_{s}\,,\, r\right)$. Since we assume a slowly varying sound speed, its constraint in this case is not significantly different from the one obtained in Sec.~\ref{intercomp}. Consequently, we must tune $c_d\simeq0.98$ so that the spectral index agrees with observations. We also find that consistency with observations demands $\epsilon_{\mathcal{D}}<0$. This resembles the result of Ref.~\cite{Khoury:2008wj}, where it was shown that DBI inflation with a decreasing sound speed results in an expanding universe, in
contrast to the case of increasing sound speed. The observables in this case $\left(n_{s}\,,\, r\right)$ are not very different from
those obtained for a constant sound speed and warp factor in Sec.~\ref{intercomp}. In fact, after an extensive numerical study we find it difficult to obtain $r\sim\mathcal{O}\left(10^{-3}\right)$ in this case. Therefore, from our analysis we conclude that DBIG inflation with a varying sound speed and constant warp factor does not bring any new features.

\subsubsection{Varying warp factor $\left(\epsilon_{f}\protect\neq0,\eta_{f}=0\right)$
and constant sound speed $\left(\epsilon_{\cd}=0\right)$}\label{varywarp}
In general, the warp factor can depend on fields not stabilised during inflation. Therefore, it is feasible to expect a time-dependent warp factor while cosmological scales are exiting the horizon. For example, in Ref.~\cite{Gmeiner:2007uw}, various solutions for warped geometries were considered in the context of DBI inflation. In the following, we consider a slowly varying warp factor in the DBI-Galileon inflation model and constrain its variation using current data. Therefore, taking $\epsilon_{f}\ll1$ we approximate the warp factor as follows
\begin{equation}
f=f_{0}\exp\left(\epsilon_{f}N\right)\simeq f_{0}\left(1+\epsilon_{f}N\right)\,,\label{warpvary}
\end{equation}
where $f_{0}$ is the initial value warp factor and $\epsilon_{f}$ is constant and treated as free parameter. Similarly to the previous
case, we set the magnitude of $f_0$ four $e$-foldings after the largest cosmological scales exit the horizon.

It is important to remark that, in contrast to the previous case, where $\lambda_{1,2}=\lambda_{1,2}(N)$ and no simple analytical solution can be found for Eq.~(\ref{varyspeddBG}), using $\epsilon_\cd=0$ and $\epsilon_f={\rm const.}$ gives $\lambda_1={\rm const.}$ and only $\lambda_{2}=\lambda_{2}(N)$. In turn, this allows us to find a simple solution to Eq.~(\ref{BGS-1}) in terms of $N$ 
\begin{equation}
H^{2}=\frac{F_{1}}{F_{3}^{2}}\exp\left(\frac{\tilde{m}^{2}N\left(2c_{\cd}^{2}(\epsilon_{f}-3)-3\epsilon_{f}+6\right)}{2c_{\cd}^{2}(c_{\cd}m_P^{2}+\tilde{m}^{2})}\right)C_{2}
+\frac{F_{2}\left(N\right)}{f_{0}F_{3}^{2}}\label{Hsol-varyf}\,,
\end{equation}
where $C_{2}$ is an integration constant, determined by the condition $\epsilon=1$ at $N_*=60$, and 
\[
F_{1}=\tilde{m}^{4}\left[2c_{\mathcal{D}}^{2}\left(\epsilon_{f}-3\right)-3\epsilon_{f}+6\right]^{2}\,,
\]
\[
F_{2}\left(N\right)=2c_{\mathcal{D}}^{2}\left(c_{\mathcal{D}}^{2}-1\right)\left\{ 2c_{\mathcal{D}}^{3}m_P^{2}\epsilon_{f}+2\tilde{m}^{2}c_{\mathcal{D}}^{2}\left[N\left(\epsilon_{f}-3\right)\epsilon_{f}+3\right]-3\left(\epsilon_{f}-2\right)\left(N\epsilon_{f}-1\right)\right\} \,,
\]
\[
F_{3}=\tilde{m}^{2}\left[2c_{\mathcal{D}}^{2}\left(\epsilon_{f}-3\right)-3\epsilon_{f}+6\right]\,.
\]

In the following we find the range of parameters $\left(c_{\mathcal{D}},\tilde{m},\epsilon_{f}\right)$ using the CMB constraints on $\left(n_{s},r\right)$. Firstly, since the sound speed is constant we obtain the same constraint as in Sec.~\ref{intercomp}, namely $c_\cd\simeq0.98$ to keep $n_s$ within its observed range.
\begin{figure}[htbp]
\centering \includegraphics[width=9cm]{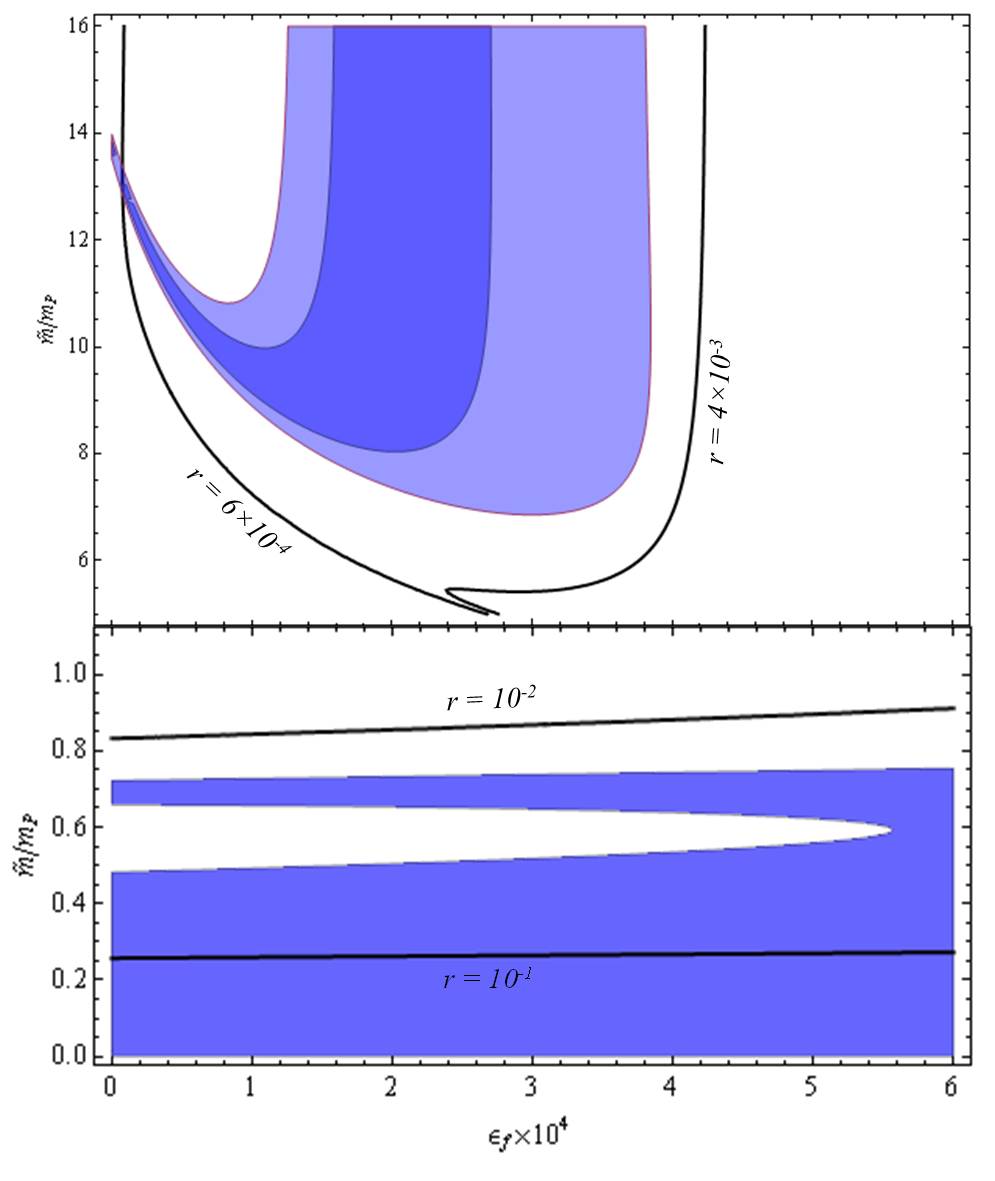}
\caption{Contour plots in the plane $\left(\tilde{m}\,,\,\epsilon_{f}\right)$. In the top panel, light and dark blue regions represent the 68\% and 95\% CL for the spectral index $n_s$, respectively. Black lines represent contours for different values of the tensor-to-scalar ratio, as indicated. In the bottom panel, the blue region depicts the 95\% CL for the spectral index $n_s$. We use $c_\cd=0.980$.}\label{varynsrntr} 
\end{figure}

In Fig.~\ref{varynsrntr} we depict the parameter space $\left(\tilde{m},\epsilon_{f}\right)$ consistent with observations of the spectral index and tensor-to-scalar ratio. Taking $c_\cd=0.98$ and enforcing \mbox{$n_s=0.968\pm0.006$}, our plot shows that it is indeed feasible to obtain a tensor-to-scalar ratio as low as \mbox{$r\simeq6\times10^{-4}$}. Nevertheless, the plot also evidences that this requires a considerable tuning between $\tilde m$ and $\epsilon_f$. We have checked that using the 2$\sigma$ interval for the spectral index does not contribute to enlarge significantly the space where $r\sim10^{-4}$. In the absence of the aforementioned tuning, expected values correspond to the range $10^{-3}\lesssim r\lesssim3\times10^{-3}$. Moreover, we have checked as well that the space where $r\sim10^{-4}$ becomes incompatible with the observed spectral index even for small deviations away from $c_\cd=0.98$. Consequently, finding $r\sim10^{-4}$ requires the combined tuning of $\tilde m,\epsilon_f$ and $c_\cd$. Nevertheless, it seems fair to say that, despite these tunings, the DBIG model of inflation represents an improvement, albeit a moderate one, with respect to the DBI model studied in Sec.~\ref{dbilimit}.

\begin{figure}[htbp]
\centering\includegraphics[width=8.0cm]{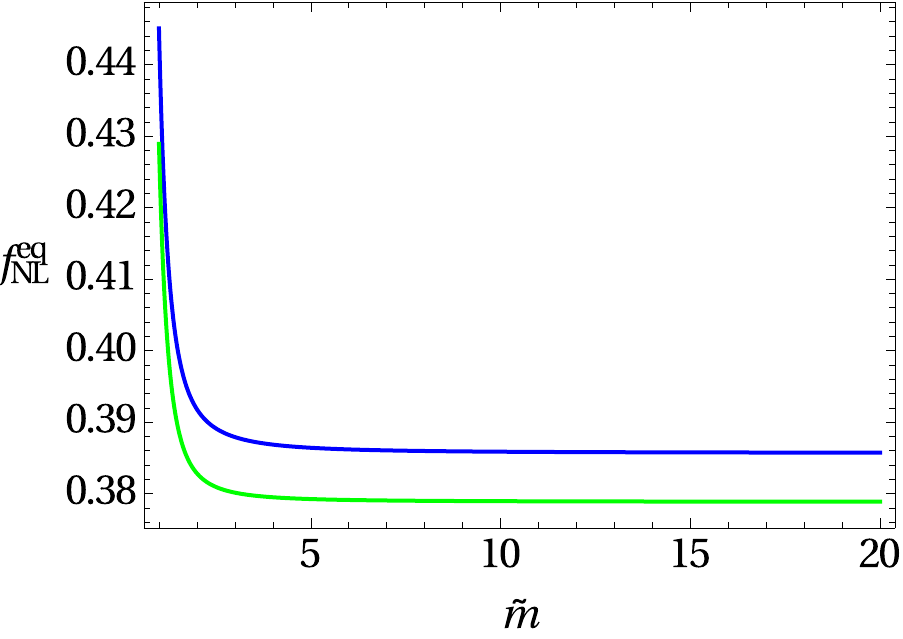}
\caption{In this plot, we depict the non-Gaussian parameter $f_{NL}^{eq}$ as a function of $\tilde{m}$ (with $\tilde m$ in units of $m_P$). We take $c_{\cd}=0.98$ and $ \epsilon_{f}\sim10^{-4} $ (Blue line) and $ \epsilon_{f}\sim10^{-6} $ (Green line). In this plot $ 0.326\leq\alpha\leq0.33 $ for $ 1\leq\tilde{m}\leq20 $.}\label{fnlvaryf} 
\end{figure}

In addition, we verify the equilateral non-Gaussianity by using the approximate expression for $f^{eq}_{NL}$ in Eq.~(\ref{fnleinter}). Since we consider a tiny variation of the warp factor we can practically neglect its contribution to non-Gaussianity. From Fig.~\ref{fnlvaryf} we can conclude that the DBIG model with varying warp factor leads to non-Gaussianities within the current observational bounds. Consequently, we conclude that after including a varying warp factor the DBIG model of inflation could be of crucial importance with respect to B-mode detection and non-Gaussianities in future CMB experiments \cite{Creminelli:2015oda}.

\begin{figure}[htbp]
\centering\includegraphics[height=2.5in]{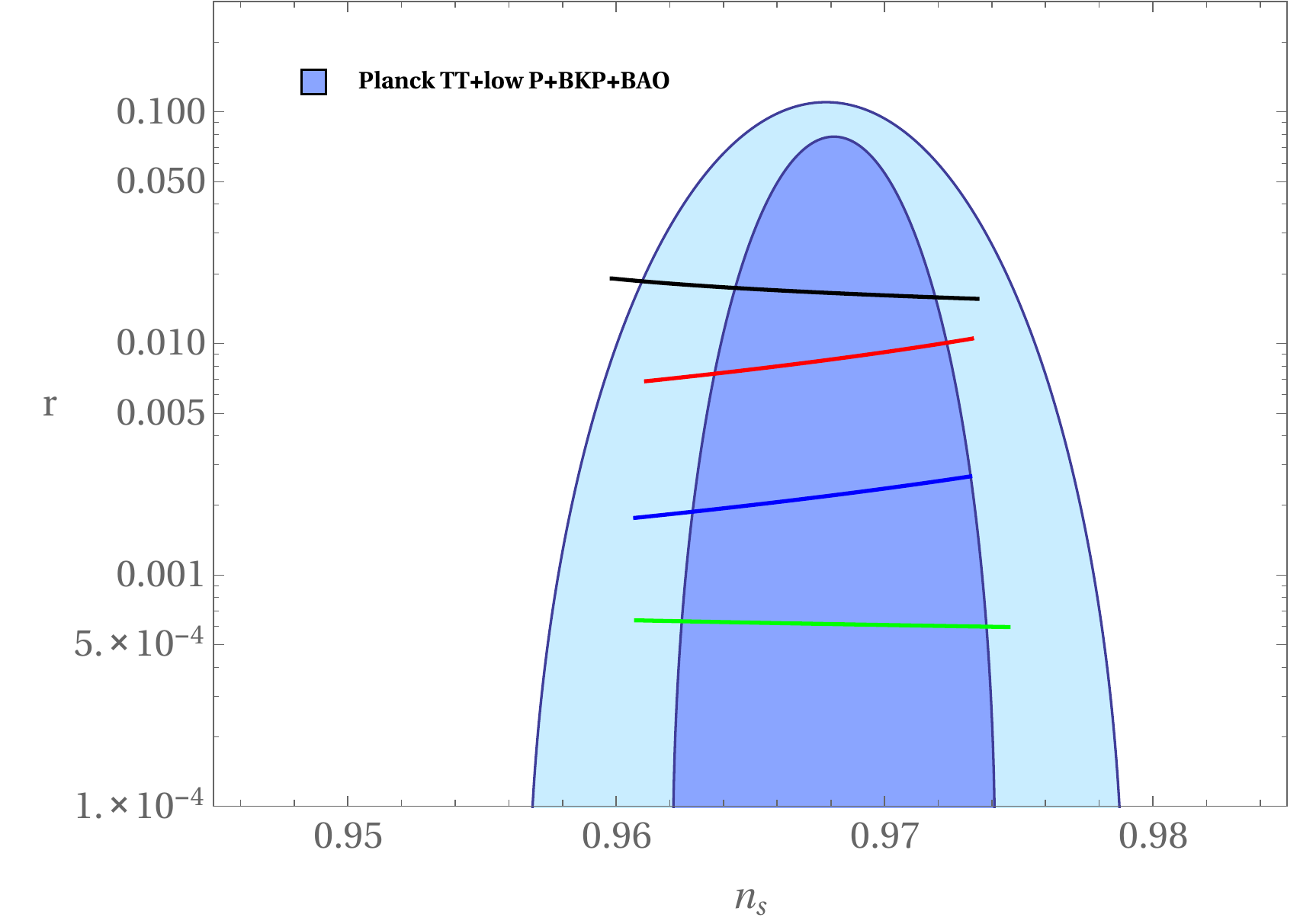}\caption{Predictions of the DBIG model for $N_*=60$ along with the Planck TT+low P+BKP+BAO constraints on the space $(n_s,r)$ at the 68\% and 95\% CL. The black line represents the case with constant sound speed and warp factor ($c_\cd=0.985$, $1\leq\tilde m/m_P\leq1.25$). Different model predictions for a constant sound speed and varying warp factor are plotted in red ($c_\cd=0.985$, $\tilde m=15m_P$ and $5.1\leq10^4\epsilon_f\leq8.5$), blue ($c_\cd=0.98$, $\tilde m=15m_P$ and $1.5\leq10^4\epsilon_f\leq2.6$) and green ($c_\cd=0.98$, $\tilde m=13m_P$ and $0.07\leq10^4\epsilon_f\leq0.11$).}\label{DBIG-total} 
\end{figure}

We finish this section by depicting the predictions of DBIG inflation for different sets of values of the model parameters in Fig.~\ref{DBIG-total} and by summarizing our results in Table~\ref{summary}. We recall that the values collected in the table were obtained taking by enforcing the scalar spectral index to lie within its observed range $n_s=0.968\pm0.006$ at the 95\% CL and taking $N_*=60$.
\begin{center}
\begin{table}[htbp]
\begin{tabular}{|c||c|c|c|c|c|}
\hline 
{\tiny{}{Inflation}}{\footnotesize{} } & {\footnotesize{}$r$ } & {\footnotesize{}$r/n_{t}$ } & {\footnotesize{}$\begin{aligned}m_{\phi}/m_P\end{aligned}
$ } & {\footnotesize{}$\begin{aligned}V_{*}^{1/4}/10^{16}\,{\rm GeV}\end{aligned}
$ } & {\footnotesize{}$f/m_P^{4}$ }\tabularnewline
\hline 
\hline 
{\tiny{}{DBI limit}}{\footnotesize{} } & {\footnotesize{}$\left(0.01,\,0.1\right)$ } & {\footnotesize{}$\left(-4.8,-0.7\right)$ } & {\footnotesize{}$6.63\times10^{-6}$ } & {\footnotesize{}$\left(0.95,1.82\right)$ } & {\footnotesize{}$\sim10^{12}-10^{14}$}\tabularnewline
\hline 
{\tiny{}{Galileon limit}}{\footnotesize{} } & {\footnotesize{}$\left(0.13,\,0.15\right)$ } & {\footnotesize{}$\left(-8.1,-7.93\right)$ } & {\footnotesize{}$m_{\phi}^{2}<0$ } & {\footnotesize{}$\left(0.64,0.70\right)$ } & {\footnotesize{}$\sim10^{9}$}\tabularnewline
\hline 
{\tiny{}{DBIG}}{\footnotesize{} } & {\footnotesize{}$\left(0.01,\,0.1\right)$ } & {\footnotesize{}$\left(-7.95,-7.5\right)$ } & {\footnotesize{}$m_{\phi}^{2}<0$ } & {\footnotesize{}$\left(1.7,2.1\right)$ } & {\footnotesize{}$\sim10^{8}-10^{9}$}\tabularnewline
\hline 
{\footnotesize{}
}{\tiny{}{(ii)Varying $f$}}{\footnotesize{} } & {\footnotesize{}$\left(0.0068,\,0.0095\right)$ } & {\footnotesize{}$\left(-7.95,\,-7.85\right)$ } & {\footnotesize{}$\left(2.41,\,2.9\right)\times10^{-7}$ } & {\footnotesize{}$\left(5.9,\,6.4\right)$ } & {\footnotesize{}$\left(6,\,9\right)\times10^{10}$}\tabularnewline
\hline 
 & {\footnotesize{}$\left(0.0018,\,0.0027\right)$ } & {\footnotesize{}$\left(-8.01,\,-7.95\right)$ } & {\footnotesize{}$\left(3.6,\,5.2\right)\times10^{-8}$ } & {\footnotesize{}$\left(4.1,\,4.6\right)$ } & {\footnotesize{}$\left(2.2\,,\,3.4\right)\times10^{9}$}\tabularnewline
\cline{2-6} 
 & {\footnotesize{}$\left(0.0006,\,0.0007\right)$ } & {\footnotesize{}$\left(-7.63,\,-7.52\right)$ } & {\footnotesize{}$\left(1.52,\,1.58\right)\times10^{-8}$ } & {\footnotesize{}$\left(2.8,\,2.85\right)$ } & {\footnotesize{}$\left(0.17,\,0.18\right)\times10^{7}$}\tabularnewline
\hline 
\end{tabular}
\caption{Inflationary observables in various limits of DBIG inflation.}\label{summary} 
\end{table}\end{center}

\section{On a class of background solutions}\label{varysols}
Until now, we have explored solutions to the background Eqs.~(\ref{Friedmann}) and (\ref{Raychaudhuri}) in which the sound speed and warp factor are either constants or time-dependent functions with very slow variation, although not simultaneously time-dependent. This choice is motivated by the simplicity of the perturbation spectrum imprinted in the CMB, which strongly favors the simplest inflationary models. Nevertheless, it is reasonable to conjecture, and to some extent expected, that in the early stages of inflation, when the observable cosmological scales are still deep within the horizon, the background dynamics has been much different from the simple slow-roll evolution supported by CMB observations. Therefore, it is interesting to investigate what kind of inflationary dynamics does the DBI-Galileon model give rise to when the  sound speed and warp factor become time-dependent functions simultaneously. In general, however, it is not possible to integrate the equations of motion for general functions $c_\cd(t)$ and $f(t)$. Owing to this difficulty, in order to find analytical solutions of the background equations we pursue a phenomenological approach in which we consider two different ansatze for the functions $\lambda_1$ and $\lambda_2$.


 


If we allow the sound speed $c_{\mathcal{D}}$ and warp factor $f$ to change ($\epsilon_{\cd},\epsilon_{f}\neq0$) the coefficients $\lambda_{1,2}$ become time-dependent functions. In such case, Eq.~(\ref{BGS-1}) can be rewritten as 
\begin{equation}
\frac{d\ln H}{\lambda_{1}-\lambda_{2}H^{-2}}=d\ln a\,.\label{csvHintegral}
\end{equation}
In what follows, we discuss two different parameterizations for $\lambda_{1,2}$ to find approximate solutions for $a(t)$.

\subsubsection*{Parametrization 1}
\begin{figure}[htbp]
\centering\includegraphics[width=4.5cm]{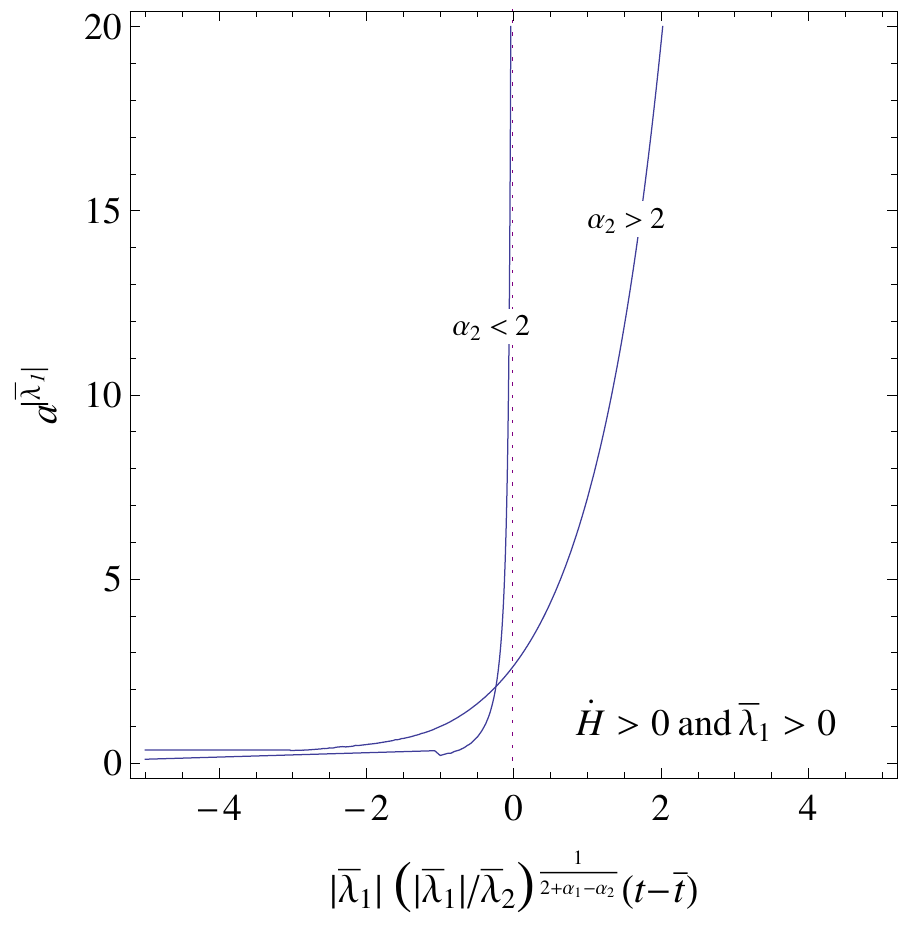}\hspace{0.25cm} \includegraphics[width=4.5cm]{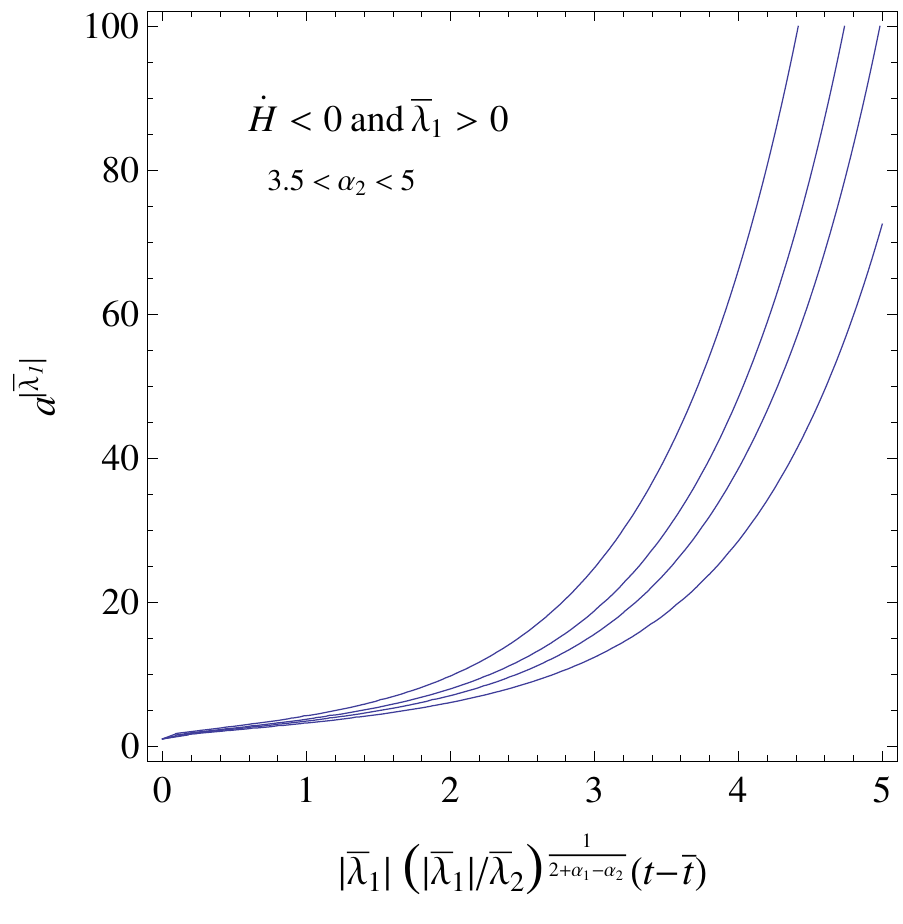}\hspace{0.25cm}\includegraphics[width=4.5cm]{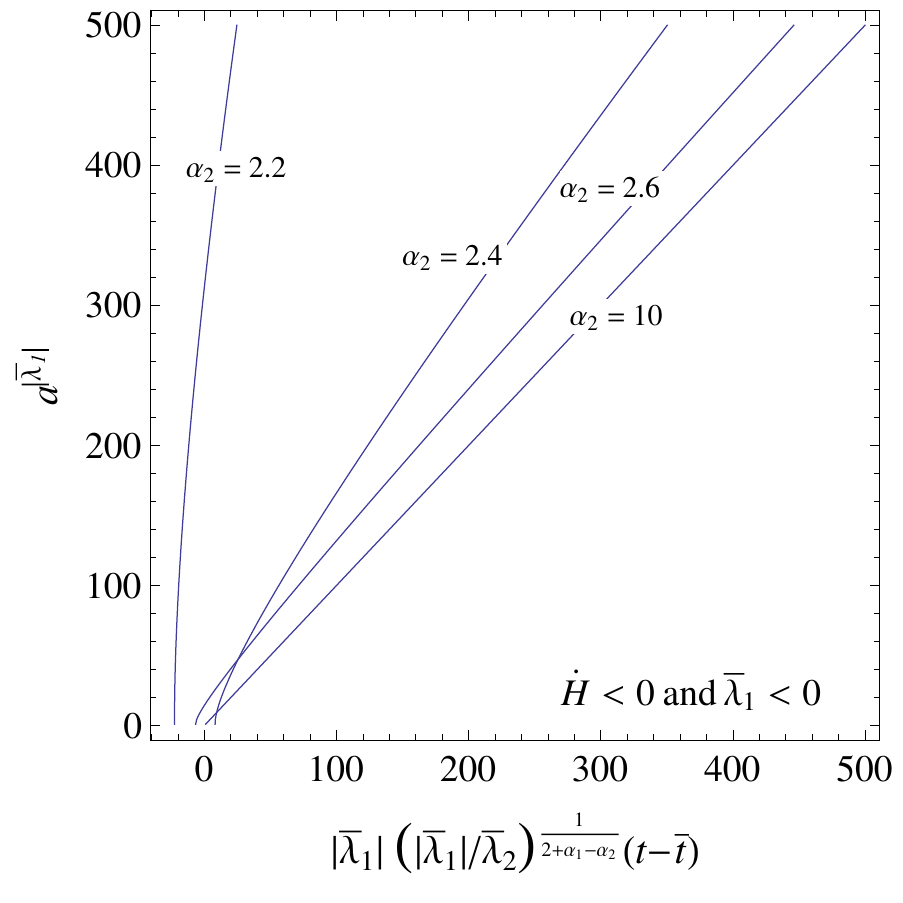}\caption{Evolution of the scale factor $a(t)$, according to Eq.~(\ref{PV1h2a}),
for $\bar{\lambda}_{1},\dot{H}>0$ (left panel), for $\bar{\lambda}_{1}>0$,
$\dot{H}<0$ (central panel) and for $\bar{\lambda}_{1},\dot{H}<0$
(right panel). For simplicity we take $\kappa=1$.}\label{fig-p1} 
\end{figure}

The simplest strategy to integrate Eq.~(\ref{csvHintegral}) is to
rewrite $\lambda_{1,2}$ as functions of $H$. Thus, we consider the
temporal dependence for $\lambda_{i}$ (with $i=1,2$) of the form
\begin{equation}
\lambda_{i}=\overline{\lambda}_{i}H^{\alpha_{i}}\,,\label{PV1}
\end{equation}
where $\overline{\lambda}_{i},\alpha_{i}$ are constants. Using this
ansatz, Eq.~(\ref{csvHintegral}) can be integrated to give 
\begin{equation}
_{2}F_{1}\left(1,1+\beta;2+\beta;\frac{\lambda_{1}H^{2}}{\lambda_{2}}\right)H^{2}=\lambda_{2}\left(\alpha_{2}-2\right)\ln|\kappa a|\quad,\quad\beta\equiv\frac{\alpha_{1}}{\alpha_{2}-\alpha_{1}-2}\,,\label{PV1hyperG}
\end{equation}
where $_{2}F_{1}$ is the hypergeometric function and $\kappa$ is
an arbitrary constant. Note that in the limit $\alpha_{1,2}\to0$
we can use the identity $_{2}F_{1}(1,1;2;z)\, z=-\ln|1-z|$ to arrive
at Eq.~(\ref{Hconscase}). Given the complexity of the above solution,
substituting $H=\dot{a}/a$ to integrate the resulting differential
equation in terms of $a(t)$ is of no practical use. Thus, it is necessary
to resort to numerical methods to integrate it. Nevertheless, if $|\beta|<1$
an approximation to the evolution equation is given by (see Appendix
\ref{ap1} for details) 
\begin{equation}
\ln\left|1-\frac{\lambda_{1}H^{2}}{\lambda_{2}}\right|\simeq\ln|\kappa a|^{A}\quad{\rm with}\quad A\equiv\frac{(2-\alpha_{2})\lambda_{1}}{1+\beta}\simeq(2-\alpha_{2})\lambda_{1}\,.\label{PV1approax}
\end{equation}
For $\alpha_{2}\lesssim{\cal O}(1)$, the condition $|\beta|\ll1$
implies $|\alpha_{1}|\ll1$. Provided $H$ does not change exponentially,
which can be certainly applied to the regular solution plotted in
Fig.~\ref{fig1}, we can approximate $\lambda_{1}$ by a constant
since \mbox{$\lambda_{1}\simeq\overline{\lambda}_{1}\left(1+\alpha_{1}\ln(H/H_{*})+\ldots\right)$}.
This reasoning can be applied to the singular solution as well whenever
it finds itself sufficiently away from the singularity at $t=\bar{t}$.
Using Eq.~(\ref{BGS-1}), we rewrite Eq.~(\ref{PV1approax}) as
\begin{equation}
H^{2+\alpha_{1}-\alpha2}=\frac{\overline{\lambda}_{2}}{|\overline{\lambda}_{1}|}\,{\rm sign}(\lambda_{1})\left(1+{\rm sign}(\dot{H})|\kappa a|^{A}\right)\,,\label{PV1h2a}
\end{equation}
which can be integrated to obtain the scale factor $a(t)$ in terms
of hypergeometric functions. The implicit function (for simplicity
we present the solution for $\kappa=1$ and vanishing $\alpha_{1}$)
which defines the scale factor is given by, 
\begin{equation}
\begin{split}\bar{\lambda}_{1}\left(t-\bar{t}\right)\approx & -\textrm{sign}\left(\dot{H}\right)\left(a^{\left(\alpha_{2}-2\right)\bar{\lambda}_{1}}+\textrm{sign}\left(\dot{H}\right)\right)\left(-\textrm{sign}\left(\bar{\lambda}_{1}\right)\frac{\bar{\lambda}_{2}\left(a^{\left(2-\alpha_{2}\right)\bar{\lambda}_{1}}+\textrm{sign}\left(\dot{H}\right)\right)}{\left|\bar{\lambda}_{1}\right|}\right){}^{\frac{1}{\alpha_{2}-2}}\\
 & _{2}F_{1}\left(1,1;1+\frac{1}{2-\alpha_{2}};-\textrm{sign}\left(\dot{H}\right)a^{\left(\alpha_{2}-2\right)\bar{\lambda}_{1}}\right)\quad,\quad\alpha_{2}\neq2
\end{split}
\label{GC-p1-sol}
\end{equation}
From Eq.~(\ref{PV1h2a}) we easily recover the background solution
with constant sound speed and constant warp factor, Eq.~(\ref{Hconscase}),
in the limit $\alpha_{1,2}\to0$. An important aspect of Eq.~(\ref{PV1h2a})
is that it only requires $|\alpha_{1}|$ to be small, whereas $|\alpha_{2}|$
can be relatively large, thus allowing a significant evolution of
$\lambda_{2}$ during inflation. Note that if we consider $c_{\mathcal{D}}$
constant, for consistency with the smallness of $\alpha_{1}$, then
from Eq.~(\ref{lambda2}) it follows that the evolution of $\lambda_{2}$
is to be attributed to the warp factor $f$. Below we study the behaviour
of the computed solution for different values of $\alpha_{2}$. In
view of Eq.~(\ref{PV1h2a}), we may consider three cases consistent
with $H^{2}>0$: 
\begin{itemize}
\item $\bar{\lambda}_{1}>0$ and $\dot{H}>0$. This case is illustrated
in the left panel of Fig.~\ref{fig-p1}, where for $\alpha_{2}<2$
we have a singular solution when $t\rightarrow\bar{t}$ . Any other
solution with $\alpha_{2}>2$ is regular at $t=\bar{t}$. 
\item $\bar{\lambda}_{1}>0$ and $\dot{H}<0$. This regime takes place provided
$(|\kappa|a)^{A}<1$. A thorough numerical study of this scenario
shows that only for a limited range of values of $\alpha_{2}$ the
integration of Eq.~(\ref{PV1h2a}) yields a well behaved physical
solution for the scale factor. In the central panel of Fig.~\ref{fig-p1}
we depict the solution for a few values of $\alpha_{2}$ in the range
$3.5<\alpha_{2}<5$ 
\item $\bar{\lambda}_{1}<0$ and $\dot{H}<0$. The constraint now is $(|\kappa|a)^{A}>1$.
This case, depicted in the right panel of Fig.~\ref{fig-p1},
possesses smooth solutions for $\alpha_{2}>2$. Moreover, for large
values of $\alpha_{2}$, the scale factor follows approximately a
power law $a\left(t\right)\sim\left(t-\bar{t}\right)^{1/\left|\bar{\lambda}_{1}\right|}$. 
\end{itemize}

\subsubsection*{Parametrization 2}
A second, simple alternative to solve Eq.~(\ref{BGS-1}) with time-dependent
$\lambda_{1,2}$ is to parametrize their dependence as 
\begin{equation}
\lambda_{i}=\lambda_{i*}(a/a_{*})^{\alpha_{i}}\,,\label{pv2}
\end{equation}
where $\lambda_{i*},\alpha_{i}$ are constants and $a_{*}$ is the
scale factor when the largest cosmological scales exit the horizon.
Defining 
\begin{equation}
z\equiv\ln H\quad,\quad y\equiv\ln(a/a_{*})\,,\label{pv2variables}
\end{equation}
we find the exact solution to Eq.~(\ref{BGS-1}) (see Appendix \ref{ap1})
\begin{equation}
e^{2z}=\frac{2\lambda_{2}}{\alpha_{1}}\left(\frac{\alpha_{1}}{2\lambda_{1}}\right)^{\frac{\alpha_{2}}{\alpha_{1}}}{\rm exp}\left(\frac{2\lambda_{1}}{\alpha_{1}}\, e^{y\alpha_{1}}\right)\,\Gamma\left(\frac{\alpha_{2}}{\alpha_{1}},\frac{2\lambda_{1}}{\alpha_{1}}\, e^{y\alpha_{1}}\right)+\kappa\,{\rm exp}\left[\frac{2\lambda_{1}}{\alpha_{1}}\left(e^{y\alpha_{1}}-1\right)\right],\label{pv2soln}
\end{equation}
where $\Gamma(s,x)$ is the incomplete Gamma function \cite{Abramowitz}.
In the limit $\alpha_{1,2}\to0$ we easily recover Eq.~(\ref{Hconscase}),
whereas for $\alpha_{1,2}\neq0$ we can use the asymptotic formula
$\Gamma(s,x)\approx x^{s-1}e^{-x}$ when $x\gg1$. In such case, the
above equation becomes 
\begin{equation}
e^{2z}\simeq\left(\frac{\lambda_{2*}}{\lambda_{1*}}\right)e^{y(\alpha_{2}-\alpha_{1})}+\kappa\exp\left[\frac{2\lambda_{1*}}{\alpha_{1}}\left(e^{y\alpha_{1}}-1\right)\right]\,.\label{pv2solution}
\end{equation}
If we focus on the background evolution while cosmological scales
are exiting the horizon then $0\leq y\lesssim9$, and $y\alpha_{1}\ll1$
provided $|\alpha_{1}|\ll1$. Neglecting higher orders in $y\alpha_{1}$
we obtain 
\begin{equation}
H^{2}\simeq\left(\frac{\lambda_{2*}}{\lambda_{1*}}\right)a^{\alpha_{2}-\alpha_{1}}+\kappa a^{2\lambda_{1*}}\,,\label{pv2h2a}
\end{equation}
which can be integrated to obtain $a(t)$ in terms of hypergeometric
functions, and also gives Eq.~(\ref{Hconscase}) in the limit $\alpha_{1,2}\to0$.
The implicit function (again, and for simplicity, we present the solution
for $\kappa=1$ and vanishing $\alpha_{1}$) that determines the scale
factor $a(t)$ is given by 
\begin{equation}
\begin{split}\left(t-\bar{t}\right)\approx- & \frac{2\lambda_{1*}a^{-\alpha_{2}}\sqrt{\frac{\lambda_{2*}a^{\alpha_{2}}}{\lambda_{1*}}+a^{2\lambda_{1*}}}\,_{2}F_{1}\left(1,\frac{\lambda_{1*}-\alpha_{2}}{2\lambda_{1*}-\alpha_{2}};\frac{-\alpha_{2}}{4\lambda_{1*}-2\alpha_{2}}+1;-\frac{a^{2\lambda_{1*}-\alpha_{2}}\lambda_{1*}}{\lambda_{2*}}\right)}{\alpha_{2}\lambda_{2*}}\,.\end{split}
\label{GC-p2-sol}
\end{equation}
Similarly to Eq.~(\ref{PV1h2a}), $|\alpha_{2}|$ is allowed to take
on relatively large values in Eq.~(\ref{pv2h2a}).

Notice that in Eq.~(\ref{GC-p2-sol}) the hypergeometric function $_{2}F_{1}$ is undefined when $\left(\frac{-\alpha_{2}}{4\lambda_{1*}-2\alpha_{2}}+1\right)$ is a negative integer. Therefore, from a formal point of view, by taking $\alpha_{2}<0$ we avoid the regions where Eq.~(\ref{GC-p2-sol}) is undefined. In addition we must also impose that $\alpha_{2}\neq2\lambda_{1*}$. When $\alpha_{2}<0$, we have from Eq.~(\ref{pv2h2a}) that $H$ becomes singular if the scale factor $a$ goes to zero. It can be checked in Eq.~(\ref{GC-p2-sol}) that $\left(t-\bar{t}\right)$ is zero whenever $a$ is zero, which amounts to have an undesirable singular solution for $H$ when $t=\bar{t}$.

In view of our results, it seems reasonable to conclude that the solutions obtained using the ansatz in Eq.~(\ref{PV1}) for $\dot{H}<0$ (with either sign of $\bar{\lambda}_{1}$) provide a more appropriate qualitative evolution for $a(t)$ than those described by the ansatz in Eq.~(\ref{pv2}). Therefore, our analysis \textit{demonstrates} that, within the context of DBI-Galileon inflation, it is possible to envisage an early inflationary stage during which the warp factor undergoes a significant variation. The relevance of this result is that such phase can be smoothly connected to the last phase of slow-roll while allowing a marginal variation of the warp factor and agreeing with current CMB observations. In this sense, it is very suggestive to imagine that the early phase of rapidly evolving geometric structure could be connected to the very beginning of inflation.

\section{Conclusions}\label{conclns}
In this paper we study the DBI-Galileon inflationary scenario, which constitutes a generic extension of the DBI model involving an induced gravity, and obtain the gravitational field equations allowing the sound speed $c_{\cd}$ and warp factor $f$ that the model depends on to be time-dependent. We find exact solutions to the background Eqs.~(\ref{Friedmann}) and (\ref{Raychaudhuri}) when $c_{\cd}$ and $f$ are constant. We obtain a singular behaviour at finite time for the scale factor and Hubble parameter when $\lambda_{2}<\lambda_{1}H^{2}$, and also a regular behaviour when $\lambda_{2}>\lambda_{1}H^{2}$ (see Fig.~\ref{fig1}). We focus on inflationary scenarios under the slow-roll approximation and constrain the model parameters using the Planck 2015 results. In addition, we constrain the warp factor in the different inflationary regimes using CMB data. Notice that the warp factor scale might be important, regarding warped string phenomenology, to understand extra dimensions and warped geometries arising from string theory.  
We find that, in general, different warped geometries give rise to distinct inflationary predictions. In the case of constant $c_{\cd}$ and $f$ (see Fig.~\ref{PspaceDBIG}), the tensor-to-scalar ratio is $r\gtrsim\mathcal{O}\left(10^{-2}\right)$. Later, we considered the DBI-Galileon model with a slowly varying warp factor and find that the tensor-to-scalar ratio can be as low as $r\simeq6\times10^{-4}$ (see Figs.~\ref{varynsrntr} and \ref{DBIG-total}). However, we find that this requires the combined tuning of $\tilde m$, $\epsilon_f$ and $c_\cd$. In any case, a varying warped geometry brings the predictions of the DBIG model closer to those of the Starobinsky model.

Another aspect of our study is the violation of the standard consistency relation of single-field inflation, $r=-8n_{t}$. Since DBIG inflation is a class of generalized G-inflation, we find deviations away from the standard consistency relation $r=-8n_{t}$. However, with the exception of the DBI limit (see Fig.~\ref{fig1-1}), the deviations found in the rest of cases under study are quite small (see Table \ref{summary}). This result is consistent with the status about the tensor consistency relation in Galileon models as it is described in Ref.~\cite{Unnikrishnan:2013rka}. We emphasize that a prominent detection of the B modes, thanks to future CMB probes devised with a greater sensitivity, can discriminate models of DBIG inflation.
 
Finally, we aim at describing an early stage of inflation taking place well before cosmological scales exit the horizon, we obtain general background solutions allowing an arbitrary time dependence for $c_{\cd}$ and $f$ by promoting the coefficients $\lambda_{1}$ and $\lambda_{2}$ in the background Eq.~(\ref{BGS-1}) to time-dependent functions. To integrate the background equations analytically we pursue a phenomenological approach, making use of the ansatze in Eqs.~(\ref{PV1}) and (\ref{pv2}). The validity of our approximations demands that $\lambda_{1}$ remains approximately constant ($\alpha_{1}\simeq0$) for both ansatze, whereas $\lambda_{2}$ can have substantial variation since $\alpha_{2}$ is not constrained to be small (see Fig.~\ref{fig-p1}). This variation of $\lambda_{2}$, in turn, can be attributed to a variation of the geometric warp factor $f$ since $c_{\cd}$ remains approximately constant. From our numerical exploration of the approximate solution we conclude that the ansatz in Eq.~(\ref{PV1}) provides a more appropriate, qualitative evolution for the scale factor. Our analysis thus provides the intriguing possibility to consider an early stage of DBI-Galileon inflation (may be even connected to its very beginning) with a significantly varying geometric structure that gives way, once the geometric structure becomes approximately stabilized, to a final phase of slow-roll in perfect agreement with current CMB observations.

\acknowledgments We thank anonymous referee for useful comments. S.K. is grateful for the support of grant SFRH/BD/51980/2012 from the Portuguese Agency Fundação para a Ciência e Tecnologia. This research work is partially supported by the grants PTDC/FIS/111032/2009 and UID/MAT/00212/2013. J.C.B.S. is supported by COLCIENCIAS grant No.110656399958. C. E-R is supported by CNPq National Fellowship and is grateful to E. Copeland for his kind hospitality and support during author's postdoc stay at Nottingham. 

\appendix

\section{Analytical approximations}

\label{ap1}

\subsection*{Parametrization 1}

Using the definition of the hypergeometric function \cite{Abramowitz}
we have 
\[\label{A1}
_{2}F_{1}\left(1,1+\beta;2+\beta;z\right)=\frac{\Gamma(2+\beta)}{\Gamma(1+\beta)}\sum_{n=0}^{\infty}\frac{\Gamma(1+n)\Gamma(1+\beta+n)}{\Gamma(2+\beta+b)}\frac{z^{n}}{n!}=(1+\beta)\sum_{n=0}^{\infty}\frac{z^{n}}{1+\beta+n}\,.
\]
For $\beta<1$ we can approximate 
\begin{equation}
\frac{1}{1+\beta+n}=\left(\frac{1}{1+n}\right)\frac{1}{1+\frac{\beta}{1+n}}\simeq\frac{1}{1+n}\left(1-\frac{\beta}{1+n}\right)=\frac{n+1-\beta}{(n+1)^{2}}\,,\label{A2}
\end{equation}
and substituting in Eq.~(\ref{A1}) we arrive at 
\begin{equation}
_{2}F_{1}\left(1,1+\beta;2+\beta;z\right)\simeq(1+\beta)\sum_{n=0}^{\infty}\frac{n+1-\beta}{(n+1)^{2}}\, z^{n}=-\frac{(1+\beta)}{z}\left(\ln|1-z|+\beta\,{\rm Li}_{2}(z)\right)\,,\label{A3}
\end{equation}
where ${\rm Li}_{n}(z)=\sum_{k=1}^{\infty}k^{-n}z^{k}$ is the polylogarithm
function \cite{Abramowitz}. Despite its being an excellent approximation
for $\beta<1$, substituting the above into Eq.~(\ref{PV1hyperG})
leads to a differential equation still too complicated (to solve for
$a(t)$) due to the polylogarithmic function ${\rm Li}_{2}(z)$. Our
aim, therefore, is to find a simple analytical solution reproducing
the qualitative behaviour of the scale factor. The simplest manner
to achieve this is to neglect the term in the polylogarithm function
in Eq.~(\ref{A3}). This simplification can be justified after approximating
\begin{equation}
\sum_{n=0}^{\infty}\frac{z^{n}}{1+\beta+n}\simeq\sum_{n=0}^{\infty}\frac{z^{n}}{1+n}=-\frac{\ln|1-z|}{z}\label{A5}
\end{equation}
in Eq.~(\ref{A1}), which holds provided $\beta\ll1$. In that case,
after substituting $z\to\lambda_{1}H^{2}/\lambda_{2}$, the resulting
background equation (Eq.~(\ref{PV1approax})) has the advantage of
being relatively simple.

\subsection*{Parametrization 2}

Using the variables $z$ and $y$ defined in Eq.~(\ref{pv2variables}),
our Eq.~(\ref{BGS-1}) becomes 
\begin{equation}
\lambda_{1*}e^{y\alpha_{1}}e^{2z}-\lambda_{2*}e^{y\alpha_{2}}-e^{2z}z'(y)=0\,.\label{A6}
\end{equation}
After multiplying by $\mu(y)=\exp\left[-(2\lambda_{1*}/\alpha_{1})e^{y\alpha_{1}}\right]$,
Eq.~(\ref{A6}) becomes an exact differential equation 
\begin{equation}
df=P(y,z)\, dy+Q(y,z)\, dz=0\,,\label{A7}
\end{equation}
where 
\begin{equation}
P(y,z)=\mu(y)\left[\lambda_{1*}e^{y\alpha_{1}}e^{2z}-\lambda_{2*}e^{y\alpha_{2}}\right]\quad\textrm{and}\quad Q(y,z)=-\mu(y)\, e^{2z}\,.\label{A8}
\end{equation}
Integral curves are of the form: $f(y,z)=\kappa$, where $\kappa$
is a constant. Integrating $f$ with respect to $y$ in the first
place we have 
\begin{equation}
f(y,z)=\int P(y,z)\, dy+g(z)\,,\label{A9}
\end{equation}
where $g(z)$ is to be computed by demanding $\partial_{z}f(y,z)=Q(y,z)$.
After integrating and solving for $g(z)$ we find that the integral
curves $f(y,z)=\kappa$ are determined by Eq.~(\ref{pv2soln}).

\section{Perturbation spectra}

\label{ap3}

\subsection{Scalar spectrum}

To quantify the amplitude and tilt of the spectrum we introduce the
variables $dy_{s}\equiv\frac{c_{s}}{a}\, dt$, $z_{s}\equiv\sqrt{2}a(\mathcal{F}_{s}\mathcal{G}_{s})^{1/4}$
and $u\equiv z_{s}\zeta$, thanks to which the action Eq.~(\ref{S2pert})
can be canonically normalized 
\begin{equation}
S_{s}^{(2)}=\frac{1}{2}\int dy_{s}\, d^{3}x\left[(u')^{2}-(\nabla u)^{2}+\frac{z_{s}''}{z_{s}}u^{2}\right]\,.\label{SS2}
\end{equation}
Imposing the flat spacetime vacuum solution in the subhorizon limit
$k/aH\to\infty$ for the perturbation mode $u$, we find we find 
\begin{equation}
u=\frac{\sqrt{\pi}}{2}\sqrt{-y_{s}}\, H_{\nu_{s}}(-ky_{s})\quad,\quad\nu_{s}^{2}-\frac{1}{4}\equiv y_{s}^{2}\frac{z_{s}''}{z_{s}}\,.\label{B2}
\end{equation}
Using now $\zeta=u/z_{s}$, we obtain the amplitude ${\cal P}_{\zeta}$
in Eq.~(\ref{pwrspectrum}).

To compute the spectral index of the scalar perturbations 
\begin{equation}
n_{s}-1\equiv3-2\nu_{s}\,,
\end{equation}
first we need to find $\nu_{s}$. Using the definition of $z_{s}$
we find 
\begin{equation}
\frac{z_{s}''}{z_{s}}=\left(\frac{Ha}{c_{s}}\right)^{2}\left[\left(1+\frac{f_{s}+g_{s}}{4}\right)^{2}+\left(1-\epsilon-\frac{f_{s}}{2}+\frac{g_{s}}{2}\right)\left(1+\frac{f_{s}+g_{s}}{4}\right)-\frac{f_{s}f_{s}^{(2)}-g_{s}g_{s}^{(2)}}{4}\right]\,.\label{B3}
\end{equation}
The next step is to integrate $dy_{s}=(c_{s}/a)\, dt$. Assuming small
and constant $\eta$ and $\eta_{s}$ to neglect second order terms
and integrating by parts we obtain 
\begin{equation}
y_{s}=-\frac{c_{s}}{(1-\epsilon-\epsilon_{s})\, aH}\left(1+\frac{\epsilon\eta+\epsilon_{s}\eta_{s}}{(\epsilon+\epsilon_{s}-1)^{2}}\right)\,,\label{B4}
\end{equation}
which is valid for somewhat large values of $\epsilon$ and $\epsilon_{s}$
provided $\eta,\eta_{s}$ are sufficiently small. Using now Eqs.~(\ref{B2})-(\ref{B4})
and neglecting second order terms in $\eta,\eta_{s}$ one arrives
at

\begin{equation}
\frac{z_{s}^{\prime\prime}}{z_{s}}=\left(\frac{aH}{c_{s}}\right)^{2}\left[\left(1+\frac{f_{s}+g_{s}}{4}\right)\left(1-\epsilon-\frac{f_{s}}{2}+\frac{g_{s}}{2}\right)+\left(1+\frac{f_{s}+g_{s}}{2}\right)^{2}-\frac{\left(f_{s}f_{s}^{(2)}-g_{s}g_{s}^{(2)}\right)}{4}\right]\label{zspp}
\end{equation}

from which one can compute the spectral index in the linear approximation
as

\begin{equation}
n_{s}-1\simeq\frac{4\epsilon_{*}+3f_{s*}-g_{s*}}{-2+2\epsilon_{*}+f_{s*}-g_{s*}}\,\,,\label{nsL-1}
\end{equation}

Using (\ref{nsL}) we can compute the running index $n'_{s}\equiv\frac{dn_{s}}{d\ln k}$.
Since we assumed $\eta,\eta_{s}$ approximately constant and small,
the use of (\ref{B4}) allows us to write 
\begin{equation}
n'_{s}=-\frac{y_{s}aH}{c_{s}}\,\frac{dn_{s}}{d\ln a}\simeq\frac{1}{(1-\epsilon-\epsilon_{s})}\left(1+\frac{\epsilon\eta+\epsilon_{s}\eta_{s}}{(\epsilon+\epsilon_{s}-1)^{2}}\right)\,\frac{dn_{s}}{d\ln a}\simeq\frac{1}{(1-\epsilon-\epsilon_{s})}\,\frac{dn_{s}}{d\ln a}\,.\label{B7}
\end{equation}
Provided $\eta$ and $\eta_{s}$ are small and approximately constant,
we can expand (\ref{nsL}) to first order in $\eta,\eta_{s}$. Using
also Eqs.~(\ref{sroolscs}) and (\ref{B7}) the running index becomes
\[
n_{s}'\simeq\frac{2\epsilon_{*}f_{s*}(4-f_{s*}+g_{s*})-2g_{s*}g_{s*}^{(2)}(1+\epsilon_{*})}{(2-2\epsilon_{*}-f_{s*}+g_{s*})^{2}}\,.
\]

\subsection{Tensor spectrum}

Similarly to the case of the scalar spectrum, we introduce the variables
$dy_{t}\equiv\frac{c_{t}}{a}\, dt$, $z_{t}\equiv\frac{a}{2}(\mathcal{F}_{t}\mathcal{G}_{t})^{1/4}$
and $u_{ij}\equiv z_{t}h_{ij}$ so that the action in Eq.~(\ref{st2})
can be canonically normalized 
\begin{equation}
S_{t}^{(2)}=\frac{1}{2}\int dy_{t}\, d^{3}x\left[(u_{ij}')^{2}-(\nabla u_{ij})^{2}+\frac{z_{t}''}{z_{t}}u_{ij}^{2}\right]\,.\label{B9}
\end{equation}
Imposing the flat spacetime vacuum solution as in Eq.~(\ref{B2})
we find 
\begin{equation}
u_{ij}=\frac{\sqrt{\pi}}{2}\sqrt{-y_{t}}\, H_{\nu_{t}}^{(1)}(-ky_{t})\, e_{ij}\,\quad,\quad\nu_{t}^{2}-\frac{1}{4}\equiv y_{t}^{2}\frac{z_{t}''}{z_{t}}\,.\label{B10}
\end{equation}
where $e_{ij}$ is the polarization tensor. Using that $h_{ij}=u_{ij}/z_{t}$
and taking into account the two polarization states, we arrive at
the expression in Eq.~(\ref{ptspectrum}).

\[
\frac{z_{t}^{\prime\prime}}{z_{t}}=\left(\frac{aH}{c_{s}}\right)^{2}\left[\left(\frac{f_{t}+g_{t}}{4}+1\right)\left(-\frac{f_{t}}{2}+\frac{g_{t}}{2}-\epsilon+1\right)+\left(\frac{f_{t}+g_{t}}{2}+1\right)^{2}-\frac{\left(f_{t}f_{t}^{(2)}-g_{t}g_{t}^{(2)}\right)}{4}\right]
\]

The spectral index of the tensor spectrum is 
\begin{equation}
n_{t}\equiv3-2\nu_{t}\,,\label{nt}
\end{equation}
If slow-roll parameters are small, the first order approximation gives
\begin{equation}
n_{t}=\frac{4\epsilon_{*}+3f_{t*}-g_{t*}}{-2+2\epsilon_{*}+f_{t*}-g_{t*}}\,.\label{nt-1}
\end{equation}

 \bibliographystyle{utphys}
\bibliography{References}

\end{document}